\documentclass[a4paper,11pt]{article}
\usepackage{jheppub_mod}
\usepackage[T1]{fontenc}

\usepackage{hyperref}
\usepackage{graphicx}
\usepackage{amsmath,amssymb,mathtools,slashed}
\usepackage{booktabs,tabulary}
\usepackage{dsfont}
\usepackage{color}
\usepackage{multirow}
\usepackage{subcaption}
\usepackage{braket}
\usepackage{adjustbox}
\usepackage{footmisc}

\newcommand{\mpi}{M_\pi}
\newcommand{\mpii}{M_{\pi^0}}

\newcommand{\beq}{\begin{equation}}
\newcommand{\eeq}{\end{equation}}
\newcommand{\diff}{\text{d}}

\newcommand{\qvec}{\mathbf{q}}
\newcommand{\kvec}{\mathbf{k}}
\newcommand{\lvec}{\mathbf{l}}

\newcommand{\Order}{\mathcal{O}}

\newcommand{\GeV}{\,\text{GeV}}

\newcommand{\Br}{\text{Br}}
\renewcommand{\Im}{\text{Im}\,}
\renewcommand{\Re}{\text{Re}\,}

\allowdisplaybreaks[1]

\title{Radiative corrections to \texorpdfstring{$\boldsymbol{\tau\to\pi\pi\nu_\tau}$}{}}

\author[a]{Gilberto Colangelo,}
\author[a]{Martina Cottini,}
\author[a]{Martin Hoferichter,}
\author[a]{and Simon Holz}

\affiliation[a]{
Albert Einstein Center for Fundamental Physics, Institute for Theoretical Physics, University of Bern, Sidlerstrasse 5, 3012 Bern, Switzerland}

\emailAdd{gilberto@itp.unibe.ch}
\emailAdd{cottini@itp.unibe.ch}
\emailAdd{hoferichter@itp.unibe.ch}
\emailAdd{holz@itp.unibe.ch}

\abstract{Hadronic $\tau$ decays present an opportunity to determine the isovector part of the hadronic-vacuum-polarization contribution to the anomalous magnetic moment of the muon in a way complementary to $e^+e^-\to\text{hadrons}$ cross sections. However, the required isospin rotation is only exact in the isospin limit, and corrections need to be under control to draw robust conclusions, most notably for $\tau\to\pi\pi\nu_\tau$ decays to determine the two-pion contribution, $a_\mu^\text{HVP, LO}[\pi\pi,\tau]$. In this work, we present a novel analysis of the required radiative corrections using dispersion relations, thereby extending in a model-independent way the previous analysis in chiral perturbation theory (ChPT) beyond the threshold region. In particular, we include the dominant structure-dependent virtual corrections from pion-pole diagrams, leading to sizable changes in the vicinity of the $\rho(770)$ resonance. Moreover, we work out the matching to ChPT and devise a strategy for a stable numerical evaluation of real-emission contributions near the two-pion threshold, which proves important to capture isospin-breaking corrections enhanced by the threshold singularity.  For the numerical analysis, we use a dispersive representation of the pion form factor including the $\rho'$, $\rho''$ resonances, perform fits to the available data sets for the $\tau\to\pi\pi\nu_\tau$ spectral function, and calculate the corresponding radiative correction factor $G_\text{EM}(s)$ in a self-consistent manner. Based on these results, we evaluate the $\tau$-specific isospin-breaking corrections to $a_\mu^\text{HVP, LO}[\pi\pi,\tau]$.}

\begin{document} 

\maketitle

\section{Introduction}
\label{sec:intro}

After the final release of the results from the Fermilab experiment,
the anomalous magnetic moment of the muon is known at an extraordinary precision of 124 parts per billion~\cite{Muong-2:2025xyk,Muong-2:2023cdq,Muong-2:2024hpx,Muong-2:2021ojo,Muong-2:2021vma,Muong-2:2021ovs,Muong-2:2021xzz,Muong-2:2006rrc}
\beq
a_\mu^\text{exp}=11\,659\,207.15(1.45)\times 10^{-10} \, ,
\eeq
which would allow for a test
of the Standard Model at an unprecedented level of $\Delta a_\mu\simeq 2\times 10^{-10}$, if the theoretical uncertainty
were commensurate with the experimental one.
Unfortunately, the current theory prediction~\cite{Aliberti:2025beg,Aoyama:2012wk,Hertzog:2025ssc,Volkov:2019phy,Volkov:2024yzc,Aoyama:2024aly,Parker:2018vye,Morel:2020dww,Fan:2022eto,Czarnecki:2002nt,Gnendiger:2013pva,Ludtke:2024ase,Hoferichter:2025yih,RBC:2018dos,Giusti:2019xct,Borsanyi:2020mff,Lehner:2020crt,Wang:2022lkq,Aubin:2022hgm,Ce:2022kxy,ExtendedTwistedMass:2022jpw,RBC:2023pvn,Kuberski:2024bcj,Boccaletti:2024guq,Spiegel:2024dec,RBC:2024fic,Djukanovic:2024cmq,ExtendedTwistedMass:2024nyi,MILC:2024ryz,FermilabLatticeHPQCD:2024ppc,Keshavarzi:2019abf,DiLuzio:2024sps,Kurz:2014wya,Colangelo:2015ama,Masjuan:2017tvw,Colangelo:2017qdm,Colangelo:2017fiz,Hoferichter:2018dmo,Hoferichter:2018kwz,Eichmann:2019tjk,Bijnens:2019ghy,Leutgeb:2019gbz,Cappiello:2019hwh,Masjuan:2020jsf,Bijnens:2020xnl,Bijnens:2021jqo,Danilkin:2021icn,Stamen:2022uqh,Leutgeb:2022lqw,Hoferichter:2023tgp,Hoferichter:2024fsj,Estrada:2024cfy,Deineka:2024mzt,Eichmann:2024glq,Bijnens:2024jgh,Hoferichter:2024vbu,Hoferichter:2024bae,Holz:2024lom,Holz:2024diw,Cappiello:2025fyf,Colangelo:2014qya,Blum:2019ugy,Chao:2021tvp,Chao:2022xzg,Blum:2023vlm,Fodor:2024jyn}
\beq
\label{amuSM}
a_\mu^\text{SM}=11\,659\,203.3(6.2)\times 10^{-10}
\eeq
falls short of this goal by about a factor of four. While recent years have seen impressive progress on many aspects of the Standard-Model prediction, especially the hadronic light-by-light (HLbL) contribution,  the precision of Eq.~\eqref{amuSM} actually deteriorated compared to the earlier report~\cite{Aoyama:2020ynm}, with a central value that shifted significantly. The origin of this development traces back to hadronic vacuum polarization (HVP), whose leading-order (LO) contribution, $a_\mu^\text{HVP, LO}$, was traditionally evaluated based on $e^+e^-\to\text{hadrons}$ cross-section measurements~\cite{Aoyama:2020ynm,Davier:2017zfy,Keshavarzi:2018mgv,Colangelo:2018mtw,Hoferichter:2019mqg,Davier:2019can,Keshavarzi:2019abf,Hoid:2020xjs}. While tensions among the $e^+e^-\to\pi^+\pi^-$ data sets~\cite{Achasov:2006vp,CMD-2:2006gxt,BaBar:2012bdw,BESIII:2015equ,KLOE-2:2017fda} existed before, after the CMD-3 measurement they increased to a level that could no longer be accommodated by a meaningful error inflation. Meanwhile, calculations in lattice QCD had sufficiently progressed that a consolidated average became possible, which, accordingly, was adopted  in Ref.~\cite{Aliberti:2025beg} for $a_\mu^\text{HVP, LO}$. However, as demonstrated by the HLbL contribution, the value of having two independent determinations for such a challenging precision test cannot be overstated, rendering a future data-driven HVP evaluation of utmost importance. To achieve this goal, a number of efforts are being pursued~\cite{Aliberti:2025beg,Colangelo:2022jxc}, among them new data for the crucial $2\pi$ channel (such as the preliminary measurement by BaBar~\cite{Polat:2026ysh}), but also improved radiative corrections and Monte-Carlo generators~\cite{Campanario:2019mjh,Ignatov:2022iou,Colangelo:2022lzg,Monnard:2021pvm,Abbiendi:2022liz,BaBar:2023xiy,Budassi:2024whw,Aliberti:2024fpq,Fang:2025mhn,Budassi:2026lmr}, as their impact could help resolve the current puzzling situation in the $e^+e^-$ measurements.

At the same time, hadronic $\tau$ decays offer a complementary avenue to determine the $2\pi$ contribution, by using an isospin rotation of $\tau^\pm\to\pi^\pm\pi^0\nu_\tau$ back to $\pi^+\pi^-$, as pioneered in Ref.~\cite{Alemany:1997tn}. In this approach, radiative corrections are even more important, as not only the radiative corrections to $\tau\to\pi\pi\nu_\tau$ have to be understood, but appropriate isospin-breaking (IB) corrections need to be applied to ensure consistency with the standard photon-inclusive definition of the LO HVP contribution as determined from $e^+e^-$ scattering, $a_\mu^\text{HVP, LO}[\pi\pi]$. To this end, first the $\tau$-specific radiative corrections need to be removed from the measured spectral function and branching fraction, then IB corrections 
to the matrix elements---the pion form factors $F_\pi^V(s)$ and $f_+(s)$ in neutral and charged channel, respectively---need to be evaluated, and in a last step the IB corrections specific to $e^+e^-$ need to be added (radiative corrections and $\rho$--$\omega$ mixing). For the last step, results extracted from the $e^+e^-\to\pi^+\pi^-$ data sets~\cite{Achasov:2006vp,CMD-2:2006gxt,BaBar:2012bdw,BESIII:2015equ,KLOE-2:2017fda,CMD-3:2023alj} in the context of dispersive analyses~\cite{Colangelo:2018mtw,Colangelo:2020lcg,Colangelo:2022prz,Hoferichter:2023sli,Stoffer:2023gba,Leplumey:2025kvv}
are available, while a number of the remaining corrections required for a full consideration of hadronic $\tau$ decays for the HVP evaluation remain quite model dependent, see Ref.~\cite{Aliberti:2025beg} for a review.

In this work, we consider various aspects of the radiative corrections to $\tau\to\pi\pi\nu_\tau$ decays~\cite{Colangelo:2025iad}, addressing a key challenge in the $\tau$-based HVP evaluation. 
After reviewing some basic formalism in Sec.~\ref{sec:formalism}, we recall the results of earlier calculations in
chiral perturbation theory (ChPT)~\cite{Cirigliano:2001er,Cirigliano:2002pv}, whose validity we aim to extend beyond the low-energy region via a dispersive analysis, see Secs.~\ref{sec:ChPT} and~\ref{sec:disp}, motivated by previous findings in the context  of $e^+e^-\to\pi^+\pi^-$~\cite{Ignatov:2022iou,Colangelo:2022lzg} that the corresponding structure-dependent virtual corrections can indeed be sizable. The matching to ChPT, as described in Sec.~\ref{sec:match}, also allows us to establish a suitable basis for the connection to short-distance (SD) contributions and lattice QCD~\cite{Bruno:2018ono}, most conveniently expressed in terms of the chiral low-energy constants (LECs). For the real-emission diagrams, we critically revisit previous calculations using resonance models~\cite{Flores-Baez:2006yiq,Flores-Baez:2007vnd,Davier:2010fmf,Miranda:2020wdg,Davier:2023fpl,Castro:2024prg}, both regarding the numerical implementation near threshold and the estimate of the resulting uncertainties, as described in Sec.~\ref{sec:ReEm}. For the numerical evaluation, we then extend the dispersive representation of the pion form factor to include the $\rho'$, $\rho''$ resonances, based on which we perform fits to the available data sets for the $\tau\to\pi\pi\nu_\tau$ spectral function 
from Belle~\cite{Belle:2008xpe}, ALEPH~\cite{ALEPH:2005qgp,Davier:2013sfa}, CLEO~\cite{CLEO:1999dln}, and OPAL~\cite{OPAL:1998rrm} in Sec.~\ref{sec:fits}. The consequences
for a $\tau$-based evaluation of the $2\pi$ contribution, $a_\mu^\text{HVP, LO}[\pi\pi,\tau]$, are discussed in Sec.~\ref{sec:amu}, before concluding in Sec.~\ref{sec:summary}. Further details on the radiative process $\tau\to\pi\pi\nu_\tau\gamma$ are relegated to App.~\ref{app:PhSpace}.


\section{Formalism}
\label{sec:formalism}

The LO HVP contribution to $a_\mu$ can be written as~\cite{Bouchiat:1961lbg,Brodsky:1967sr}
\begin{equation}
    a_\mu^\text{HVP, LO}=\bigg(\frac{\alpha m_\mu}{3\pi}\bigg)^2\int_{s_{\text{thr}}}^\infty\frac{\text{d}s}{s^2}\hat{K}(s)R_{\text{had}}(s)\,,
    \label{eq:amu}
\end{equation}
where the kernel function is
\begin{align}
    \hat{K}(s)&=\frac{3s}{m_\mu^2}\left[\frac{x^2}{2}(2-x^2)+\frac{(1+x^2)(1+x)^2}{x^2}\left(\log{(1+x)}-x+\frac{x^2}{2}\right)+\frac{1+x}{1-x}x^2\log{x}\right]\,,\notag\\
   x&=\frac{1-\sigma_\mu(s)}{1+\sigma_\mu(s)}\,,\qquad 
   \sigma_\ell(s)=\sqrt{1-\frac{4m_\ell^2}{s}}\,,
    \label{eq:K}
\end{align}
and $R_{\text{had}}(s)$ is related to the hadronic cross section by
\begin{equation}
    R_{\text{had}}(s)=\frac{3s}{4\pi\alpha^2}\frac{s\sigma_e(s)}{s+2m_e^2}\;\sigma\big[e^+e^-\rightarrow \text{hadrons}(\gamma)\big](s)\, .\label{eq:R}
\end{equation}
Crucially, it is defined photon-inclusively, meaning that the threshold becomes $s_\text{thr}=\mpii^2$ due to the $\pi^0\gamma$ channel and that final-state radiation (FSR) is included, while vacuum-polarization corrections are removed to avoid double counting with higher-order HVP iterations~\cite{Calmet:1976kd,Kurz:2014wya,Hoferichter:2021wyj}. 
Among all possible hadronic final states, the contribution of the two-pion intermediate state accounts for more then $70\%$ of the total LO effect and can be expressed in terms of the pion vector form factor (VFF) $F_\pi^V(s)$, describing the matrix element of the electromagnetic current
\begin{align}
    \langle\pi^\pm(p')|j_\text{em}^\mu(0)|\pi^\pm(p)\rangle&=\pm(p'+p)^\mu F_\pi^V[(p'-p)^2]\, ,\notag\\
    j_\text{em}^\mu&=\frac{1}{3}\Big(2\bar u\gamma^\mu u-\bar d\gamma^\mu d-\bar s\gamma^\mu s\Big)\,,
    \label{eq:PVFF}
\end{align}
according to
\begin{equation}
    \sigma\big[e^+e^-\rightarrow\pi^+\pi^-\big](s)=\frac{\pi\alpha^2}{3s}\big[\sigma_\pi(s)\big]^3\big|F_\pi^V(s)\big|^2\frac{s+2m_e^2}{s\sigma_e(s)}\, .
\end{equation}
Thanks to 
a conserved-vector-current (CVC) relation between electromagnetic and weak form factors, in the isospin limit, the purely hadronic cross section for $e^+ e^- \to \pi^+ \pi^-$ can be related to the $\tau^- \to \pi^- \pi^0 \nu_\tau$ differential decay width in this limit by
\begin{equation}
    \sigma\big[e^+ e^- \to \pi^+ \pi^-\big] (s) = \frac{1}{\mathcal{N}(s) \Gamma_e^{(0)}} \frac{\text{d} \Gamma \big[\tau^\pm \to \pi^\pm \pi^0 \nu_\tau\big]}{\text{d} s}\, ,
\end{equation}
where constants and phase-space factors are collected in
\begin{equation}
    \mathcal{N}(s) = \frac{3 |V_{ud}|^2}{2 \pi \alpha^2 m_\tau^2} s \left(1 - \frac{s}{m_\tau^2} \right)^2 \left(1 +
    \frac{2s}{m_\tau} \right) \, ,
\end{equation}
and further (electroweak) constants appear in
\begin{equation}
\label{Gamma_e_LO}
    \Gamma_e^{(0)} = \frac{G_F^2 m_\tau^5}{192 \pi^3} \, ,
\end{equation}
which contains the Fermi constant $G_F$~\cite{MuLan:2012sih}.
Including IB to LO, $\mathcal{O}\big(m_u-m_d,e^2\big)$, the modified CVC relation takes the form
\begin{equation}
    \sigma\big[e^+ e^- \to \pi^+ \pi^-(\gamma)\big] (s) = \frac{1}{\mathcal{N}(s) \Gamma_e} \frac{\text{d} \Gamma \big[\tau^\pm \to \pi^\pm \pi^0 \nu_\tau(\gamma)\big]}{\text{d} s}\frac{R_{\text{IB}}(s)}{S_{\text{EW}}^{\pi\pi}}\, ,
    \label{eq:cross_sec_IB}
\end{equation}
where $S_\text{EW}^{\pi\pi}=1 + 2 \alpha/\pi\log (M_Z/m_\tau)    + \cdots=1.0233(3)(24)$~\cite{Sirlin:1981ie,Marciano:1985pd,Marciano:1988vm,Marciano:1993sh,Braaten:1990ef,Erler:2002mv,Davier:2002dy,Cirigliano:2023fnz} takes into account the SD electroweak corrections in the convention that the normalization proceeds with respect to the full decay rate $\Gamma_e\equiv\Gamma_e[\tau\to e\nu_\tau\bar\nu_e]$, of which Eq.~\eqref{Gamma_e_LO} represents the LO approximation.  The IB effects are included via
\begin{equation}\label{eq:RIB}
    R_{\text{IB}}(s)=\frac{\text{FSR}(s)}{G_{\text{EM}}(s)}\frac{\big[\beta_{\pi\pi}(s)\big]^3}{\big[\beta_{\pi\pi^0}(s)\big]^3}\Bigg|\frac{F_\pi^V(s)}{f_+(s)}\Bigg|^2\, ,
\end{equation}
where $\text{FSR}(s)=1+\frac{\alpha}{\pi} \eta(s)$~\cite{Hoefer:2001mx,Czyz:2004rj,Gluza:2002ui,Bystritskiy:2005ib} accounts for final-state radiation in $e^+e^-\to\pi^+\pi^-(\gamma)$, $G_\text{EM}(s)$ for the analog radiative corrections in $\tau^\pm\to\pi^\pm\pi^0\nu_\tau(\gamma)$, and the phase-space factors are expressed in terms of 
\begin{equation}
    \beta_{\pi\pi^0}(s)=\lambda^{1/2}\bigg(1,\frac{\mpi^2}{s},\frac{\mpii^2}{s}\bigg)\,,\qquad \lambda(a,b,c)=a^2+b^2+c^2-2(ab+ac+bc)\,.
\end{equation}
Here and throughout we follow the convention that a pion without index is identified with a charged pion, in particular, $\mpi\equiv M_{\pi^\pm}$ and thus $\beta_{\pi\pi}(s)=\sigma_\pi(s)$. The final factor in Eq.~\eqref{eq:RIB} refers to IB in the matrix elements, i.e., the pion VFF as measured in $e^+e^-$ scattering, $F_\pi^V(s)$, or the weak decay, $f_+(s)$. In the latter case, in principle, a second form factor $\tilde f_-(s,t)$ contributes to the matrix element, see Eq.~\eqref{Eq:Mgen} below, but only at second order in IB. At LO in IB, the $\tau$ decay rate can therefore be expressed as 
\begin{align}
\frac{\text{d}\Gamma[\tau\to\pi\pi\nu_\tau(\gamma)]}{\text{d}s}&=S_\text{EW}^{\pi\pi}K_\Gamma(s)\big[\beta_{\pi\pi^0}(s)\big]^3|f_+(s)|^2 G_\text{EM}(s)\,,\notag\\
K_\Gamma(s)&= \frac{\Gamma_e \vert V_{ud}\vert^2}{2 m_\tau^2} \bigg(1-\frac{s}{m_\tau^2}\bigg)^2\bigg(1+\frac{2s}{m_\tau^2}\bigg)\,.
\label{decay_rate}
\end{align}
$G_{\text{EM}}(s)$ constitutes the central object of this work. Following Refs.~\cite{Cirigliano:2001er,Cirigliano:2002pv}, it can be defined as
\begin{equation}\label{eq:GEM}
    G_{\text{EM}}(s)=\frac{\int_{t_\text{min}(s)}^{t_\text{max}(s)}\text{d}t\; D(s,t)\Delta(s,t)}{\int_{t_\text{min}(s)}^{t_\text{max}(s)}\text{d}t \;D(s,t)}\, ,
\end{equation}
with integration boundaries
\beq
\label{tminmax}
t_{\text{min}/\text{max}}(s)=\mpii^2
+\frac{m_\tau^2-s}{2s}\Big(s-\mpi^2+\mpii^2\mp s\beta_{\pi\pi^0}(s)\Big)
\eeq
and tree-level kinematic function
\begin{equation}
\label{eq:Dst}
    D(s,t)=\frac{m_\tau^2}{2}\left(m_\tau^2-s\right)+2\mpi^2\mpii^2-2t\left(m_\tau^2-s+\mpi^2+\mpii^2\right)+2t^2\, .
\end{equation}
The remaining integrand 
\begin{equation}
    \Delta(s,t)=1+2\,\Re \tilde f_+(s,t)+g_\text{rad}(s,t)\,,\qquad g_\text{rad}(s,t)=g_{\text{Low}}(s,t)+g_{\text{rest}}(s,t)\,,
\label{eq:delta}
\end{equation}
contains both virtual and real photon corrections, $\tilde f_+(s,t)$ and $g_\text{rad}(s,t)$, respectively (see below for their definition), where the latter is further separated into the leading contribution from Low's theorem~\cite{Low:1958sn} and the rest.   

In practice, we do not attempt a full calculation of second-order IB effects, otherwise, also the form factor $\tilde f_-(s,t)$ would need to be kept, the kinematics in the loop integrals would become significantly more complicated, and one would have to distinguish between $F_\pi^V(s)$ and $f_+(s)$ even in the radiative corrections. In general, however, such second-order corrections are tiny, and we therefore evaluate $G_\text{EM}(s)$ in the isospin limit, $\mpii\to\mpi$, $\tilde f_-(s,t)\to 0$, $F_\pi^V(s)\to f_+(s)$. The only exception where the pion mass difference does matter in the end concerns the HVP integral, as otherwise fake IB effects can be generated if the phase-space boundaries do not match, i.e., the threshold singularity in $G_\text{EM}(s)$ sits at $s=4\mpi^2$, but the physical threshold of $\tau^\pm\to\pi^\pm\pi^0\nu_\tau$ at $s=(\mpi+\mpii)^2$. This mismatch can be avoided by a simple linear mapping~\cite{Colangelo:2018jxw}  
\begin{equation}
    G_\text{EM}(s)\mapsto G_\text{EM}\big[\tilde{s}(s)\big]\,,
\end{equation}
with
\begin{equation}\label{eq:smap}
    \tilde{s}(s) = \frac{(m_\tau^2-4M_\pi^2)s+\big[4M_\pi^2-(M_\pi+M_{\pi^0})^2\big]m_\tau^2}{m_\tau^2-(M_\pi+M_{\pi^0})^2}\, ,
\end{equation}
fulfilling $\tilde{s}[(\mpi+\mpii)^2]=4\mpi^2$ and $\tilde{s}(m_\tau^2)=m_\tau^2$, to ensure that the singularity 
in $G_\text{EM}\big[\tilde{s}(s)\big]$ is shifted to the physical threshold. In fact, this threshold singularity leads to the only second-order IB effects that do need to be included, i.e., the threshold-enhanced terms in $G_\text{EM}(s)$ multiplied with the IB corrections from phase space and $S_\text{EW}^{\pi\pi}$ lead to nonnegligible contributions, so that a linearized form of the corrections in Eqs.~\eqref{eq:cross_sec_IB} and~\eqref{eq:RIB} should be avoided, see Sec.~\ref{sec:amu}. 

\section{ChPT representation}
\label{sec:ChPT}

At LO in the chiral power counting, the $SU(3)$ ChPT Lagrangian including virtual photons and leptons and relevant for calculating the radiative corrections to $\tau^- \to \pi^- \pi^0 \nu_\tau$ involves the terms
\begin{align}\label{Eq:LOLag}
    \mathcal{L}_\text{eff}^\text{LO} \supset\ &\left[\left(\partial_\mu - i e A_\mu\right)\pi^-\right] \left[\left(\partial^\mu + i e A^\mu\right)\pi^+\right] +\bar{\tau} \left[i\gamma_\mu \left(\partial^\mu - i e A^\mu\right) - m_\tau\right] \tau \notag \\
    &+ i \bar{\nu}_{\tau \text{L}} \slashed{\partial} \nu_{\tau \text{L}} + 2 i G_F V_{ud}^*\, \bar{\nu}_{\tau \text{L}} \gamma^\mu \tau \left[\pi^0 \left(\partial_\mu + i e A_\mu\right) \pi^+ - \pi^+ \partial_\mu \pi^0 \right] \, ,
\end{align}
where $e=\sqrt{4\pi\alpha}$ is the elementary charge, $G_F$ the Fermi constant, and $V_{ud}$ the CKM matrix element.
By means of the interactions contained in the LO Lagrangian, the tree-level amplitude for $\tau^-(l_1) \to \pi^-(q_1) \pi^0 (q_2) \nu_\tau (l_2)$ is found to be
\begin{equation}
\label{ChPT_tree}
    \mathcal{M}_\text{tree} = G_F V_{ud}^* \bar{u}(l_2, \nu_\tau) (1+\gamma_5) (\slashed{q}_2 - \slashed{q}_1) u(l_1,\tau) \, .
\end{equation}
The relevant counterterms of the radiative corrections to the hadronic $\tau$ decay originate from terms in the next-to-leading-order (NLO) Lagrangians $\mathcal{L}_{e^2p^2}$ and $\mathcal{L}_\text{lept}$~\cite{Urech:1994hd,Knecht:1999ag}.

\subsection{General considerations}

In general, the decay amplitude of $\tau^-(l_1) \to \pi^-(q_1) \pi^0 (q_2) \nu_\tau (l_2)$ can be written in terms of two form factors by
\begin{align}
\label{Eq:Mgen}
    \mathcal{M} &= G_F V_{ud}^*f_+(s) \bar{u}(l_2,\nu_\tau) \gamma^\mu (1 - \gamma_5) u(l_1,\tau)\notag\\
    &\qquad\times\left[(q_2 - q_1)_\mu \big(1+\tilde f_+(s,t)\big) + (q_1 + q_2)_\mu \tilde f_-(s,t)\right] \, ,
\end{align}
where $1+\tilde f_+(s,t)$ represents the contribution due to the $J^P=1^-$ component of the weak current and $\tilde f_-(s,t)$ vanishes in the isospin limit.
We have factored out $f_+(s)$, which is defined in analogy to Eq.~\eqref{eq:PVFF}
\begin{align}
 \langle\pi^\pm(p')|j_{W^\mp}^\mu(0)|\pi^0(p)\rangle&=\mp \sqrt{2}\Big((p'+p)^\mu f_+[(p'-p)^2]+(p-p')^\mu f_-[(p'-p)^2]\Big)\, ,\notag\\
 j_{W^-}^\mu&=\bar u\gamma^\mu d\,,\qquad 
 j_{W^+}^\mu=\bar d\gamma^\mu u\,,
    \label{eq:PVFF_weak}
\end{align}
normalized in such a way that $f_+(s)=F_\pi^V(s)$ in the isospin limit, and also separated the $\Order(e^2)$ corrections represented by $\tilde f_+(s,t)$. In these conventions, $\tilde f_+(s,t)\to 0$ in the limit in which the matrix element~\eqref{Eq:Mgen} is mediated by a single current insertion. In principle, Eq.~\eqref{eq:PVFF_weak} also involves a second form factor $f_-(s)$, which is related to $\tilde f_-(s,t)$ in the same single-current approximation and constitutes an example of a second-class current~\cite{Weinberg:1958ut,Descotes-Genon:2014tla}. As shown in Ref.~\cite{Descotes-Genon:2014tla}, $f_-(s)$ can be traded for a scalar form factor $f_0(s)$, with $f_0(0)=1$ and $f_-(s)=\frac{\Delta_\pi}{s}\big[f_+(s)-f_0(s)\big]$, which shows that $f_-(s)$ scales with the pion mass difference $\Delta_\pi=\mpi^2-\mpii^2$.

Once radiative corrections are included, the form factors $\tilde f_\pm(s,t)$ do depend on a second Mandelstam variable $t$, defined as
\begin{align}
    s &= (l_1 - l_2)^2 = (q_1 + q_2)^2\, ,\notag \\
    t &= (l_1 - q_1)^2 = (q_2 + l_2)^2\, , \notag \\
    u &= (l_1 - q_2)^2 = (q_1 + l_2)^2\, , 
\end{align}
with $s+t+u=m_\tau^2+\mpi^2+\mpii^2$.
Projecting Eq.~\eqref{ChPT_tree} onto the form factors therefore implies
\begin{equation}
    \tilde f_+^\text{tree}(s,t) = \tilde f_-^\text{tree}(s,t) = 0 \, ,\qquad f_+^\text{tree}(s)=1\,.
\end{equation}
The spin-averaged squared amplitude can be expressed by
\begin{align}
    \frac{1}{2}\sum\limits_\text{spins}|\mathcal{M}|^2&= 2 G_F^2 |V_{ud}|^2|f_+(s)|^2 \Big\lbrace m_\tau^2(m_\tau^2 - s) |\tilde f_-(s,t)|^2 \\
    & -  2 m_\tau^2 (2M_{\pi^0}^2+m_\tau^2-s-2t) \Re \left[\big(1+\tilde f_+(s,t)\big) \tilde f_-^*(s,t)\right]\nonumber \\
    &+  [ 4 M_\pi^2 (M_{\pi^0}^2 - t) + 4t (s+t-M_{\pi^0}^2) -m_\tau^2 (s+4t) +m_\tau^4 ] \big|1+\tilde f_+(s,t)\big|^2 \Big\rbrace \, .\notag
\end{align}
Working at LO in IB, we use the formula above with $\tilde{f}_-(s,t)\to 0$ and $\big|1+\tilde{f}_+(s,t)\big|^2\to 1+2\, \Re\tilde{f}_+(s,t)$ in the following.
Furthermore, the differential decay width is given by
\begin{equation}
    \text{d} \Gamma = \frac{1}{32 m_\tau^3 (2\pi)^3} \frac{1}{2}\sum\limits_{\text{spins}} \left| \mathcal{M} \right|^2 \text{d}s\, \text{d}t \, ,
\end{equation}
with the phase space in the $\pi^0 \nu_\tau$ invariant mass squared limited by $t_\text{min}(s) \le t \le t_\text{max}(s)$, see Eq.~\eqref{tminmax},  and  in the $\pi^- \pi^0$ invariant mass squared  by
$(M_\pi+M_{\pi^0})^2 \le s \le m_\tau^2$. Reducing Eq.~\eqref{Eq:Mgen} to the single-current-insertion limit $f_+(s) \tilde f_-(s,t) \to f_-(s)$, $\tilde f_+(s,t)\to 0$, see Eq.~\eqref{eq:PVFF_weak}, the integral over $t$ reproduces~\cite{Cirigliano:2001er}
\begin{align}
\frac{\text{d}\Gamma[\tau\to\pi\pi\nu_\tau]}{\text{d}s}&=\frac{\Gamma_e^{(0)}|V_{ud}|^2}{2m_\tau^2}\beta_{\pi\pi^0}(s)\bigg(1-\frac{s}{m_\tau^2}\bigg)^2\bigg[\bigg(\big[\beta_{\pi\pi^0}(s)\big]^2\Big(1+\frac{2s}{m_\tau^2}\Big)+\frac{3\Delta_\pi^2}{s^2}\bigg)\big|f_+(s)\big|^2\notag\\
&-\frac{6\Delta_\pi}{s}\Re\big[f_+(s)f_-^*(s)\big]+3\big|f_-(s)\big|^2\bigg]\,.
\end{align}

\subsection{Radiative corrections}

\begin{figure}[t]
    \centering
    \includegraphics[width=0.8\linewidth]{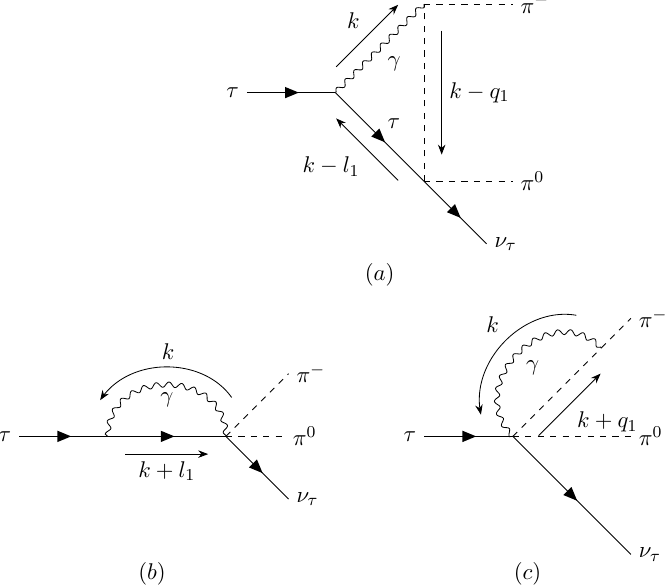}
    \caption{Leading diagrams for the radiative corrections to $\tau^- (l_1) \to \pi^-(q_1) \pi^0(q_2) \nu_\tau(l_2)$, excluding wave-function renormalization.}
    \label{fig:loops}
\end{figure}

At $\mathcal{O}(e^2p^2)$ in ChPT, the diagrams arising from the LO effective Lagrangian of Ref.~\cite{Knecht:1999ag} are shown in Fig.~\ref{fig:loops}. In this work, we employ dimensional regularization to deal with both the UV and the IR divergences arising in the computation. Following the definition in Eq.~\eqref{Eq:Mgen}, the form factors $\tilde f_+$ for the diagrams in Fig.~\ref{fig:loops} can be expressed as 
\begin{align}
\label{eq:fpBoxChpt}
    \tilde f_+^{(a)}(s,t)&=\frac{e^2}{16\pi^2}
    \Bigg\{2\big(\mpi^2+m_\tau^2-t\big)\bigg[C_0^{\pi\tau}(t)-\frac{t B_0^{\pi\tau}(t)}{\lambda_{\pi\tau}(t)}\Big(32\pi^2\Lambda_\text{IR}+1-\log\frac{\mu_\text{IR}^2}{\mpi m_\tau}\Big)\bigg]\notag\\
    &+\frac{m_\tau^2}{t}\log\frac{\mpi}{m_\tau}-\frac{2t^2-t(m_\tau^2+2\mpi^2)-m_\tau^2(m_\tau^2-\mpi^2)}{\lambda_{\pi\tau}(t)}B_0^{\pi\tau}(t)\notag\\
    &+2+2\log\frac{\mpi}{m_\tau}-\frac{7}{4}\Big(32\pi^2\Lambda_\text{UV}-\log\frac{\mu_\text{UV}^2}{\mpi^2}\Big)
    \Bigg\}\, , \notag\\
    \tilde f_+^{(b)}(s,t)&=0 \, , \notag\\
    \tilde f_+^{(c)}(s,t)&=\frac{e^2}{64\pi^2}\Big[-4+3\Big(32\pi^2\Lambda_\text{UV}-\log\frac{\mu_\text{UV}^2}{\mpi^2}\Big)\Big]\, .
\end{align}
To facilitate the comparison to the dispersive calculation later, we recast the scalar loop integrals, reduced via standard Passarino--Veltman techniques~\cite{Passarino:1978jh,Consoli:1979xw,Veltman:1980fk,Green:1980bd,tHooft:1978jhc,Melrose:1965kb,Denner:1991kt}, in the form as implemented in FeynCalc~\cite{Mertig:1990an,Shtabovenko:2016sxi,Shtabovenko:2020gxv,Shtabovenko:2023idz} and Package X~\cite{Patel:2015tea,Patel:2016fam}, with conventions translated as described, e.g., in Ref.~\cite{Shtabovenko:2016whf}. In particular, the ChPT loop diagrams in Eq.~\eqref{eq:fpBoxChpt} are most conveniently written in terms of~\cite{Ellis:2007qk,Gogniat:2025eom} 
\begin{align} 
B_0^{\pi\tau}(t) &\equiv \texttt{DiscB}[t,M_\pi,m_\tau]
=\frac{\lambda^{1/2}_{\pi\tau}(t)}{t}\log x_t\,,\notag\\ 
C_0^{\pi\tau}(t) &\equiv \texttt{ScalarC0IR6}[t,\mpi,m_\tau]
=\frac{1}{\lambda_{\pi\tau}^{1/2}(t)}\bigg[\frac{1}{2}\log^2 x_t-\frac{1}{2}\log^2\frac{\mpi}{m_\tau}\notag\\
&+\text{Li}_2\big(1-x_t^2\big)-\text{Li}_2\Big(1-x_t\frac{\mpi}{m_\tau}\Big)-\text{Li}_2\Big(1-x_t\frac{m_\tau}{\mpi}\Big)\bigg]\,,\notag\\
x_t&=-\frac{t-(m_\tau-\mpi)^2-\lambda^{1/2}_{\pi\tau}(t)}{t-(m_\tau-\mpi)^2+\lambda^{1/2}_{\pi\tau}(t)}\,,\qquad \lambda_{\pi\tau}(t)=\lambda(t,\mpi^2,m_\tau^2)\,. \label{Eq:B0_C0_lambda}
\end{align}
At this order in ChPT, the form factor $\tilde f_+(s,t)$ only depends on the Mandelstam variable $t$, while $\tilde f_-(s,t)$ vanishes in the isospin limit, and will therefore not be discussed anymore in the following.  The divergences are collected in the factors
\beq
 \Lambda= \frac{\mu^{d-4}}{16 \pi^2} \left\{ \frac{1}{d-4} - \frac{1}{2} \big[ \log (4 \pi) - \gamma_E + 1\big] \right\} \, ,   
\eeq
where the subscript differentiates between UV and IR divergences. In both cases we include the tadpole to be consistent with the typical ChPT convention~\cite{Gasser:1983yg,Gasser:1984gg}.

\begin{figure}[t]
    \centering
    \includegraphics[width=0.3\linewidth]{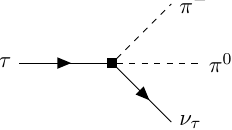}
    \caption{Counterterm diagram for the radiative corrections to $\tau^- \to \pi^- \pi^0 \nu_\tau$.}
    \label{Fig:taudec_counter}
\end{figure}

\begin{figure}[tb]
    \centering
    \includegraphics[width=\linewidth]{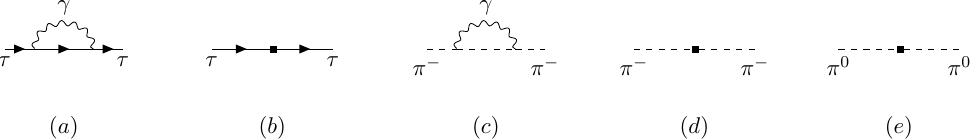}
    \caption{Loop and counterterm diagrams of the external-leg contributions.}
    \label{Fig:ext_legs}
\end{figure}

In addition to the one-loop diagrams in Fig.~\ref{fig:loops}, the full calculation includes counterterms (CT), see Fig.~\ref{Fig:taudec_counter}, as well as the self-energy (SE) corrections shown in Fig.~\ref{Fig:ext_legs}. These contributions combine to 
\begin{align}
\label{CT+SE}
    \tilde f_+^{\text{CT}+\text{SE}}(s,t)&=\frac{e^2}{16\pi^2}\bigg[64\pi^2\Lambda_\text{IR}-2\log\frac{\mu_\text{IR}^2}{\mpi m_\tau}+32\pi^2\Lambda_\text{UV}+\frac{1}{2}\log\frac{\mu_\text{UV}^2}{\mpi^2}\notag\\
    &-8\pi^2\Big(X_6^r(\mu_\text{UV})-4K_{12}^r(\mu_\text{UV})+\frac{4}{3}X_1\Big)-\frac{1}{2}-\log\frac{\mpi}{m_\tau}\bigg]\,,
\end{align}
where we already separated the UV-divergent parts of the LECs according to the $\beta$ functions from Refs.~\cite{Urech:1994hd,Knecht:1999ag}. One can easily check that the sum of Eqs.~\eqref{eq:fpBoxChpt} and~\eqref{CT+SE} is indeed UV finite. IR divergences remain, but will be canceled by real-emission diagrams, to which we return in Sec.~\ref{sec:ReEm}. The ChPT result in the form~\eqref{eq:fpBoxChpt} serves as convenient point of reference for the dispersive calculation, since $\tilde f_+^{(a)}(s,t)$ needs to be reproduced in the appropriate limit.

\section{Dispersive analysis}
\label{sec:disp}

As a first modification beyond the pure ChPT analysis summarized in Sec.~\ref{sec:ChPT}, it is clear that the strong final-state interactions between $\pi^-$ and $\pi^0$ need to be taken into account, after all, the main purpose of studying this decay concerns a precision measurement of the corresponding pion VFF. This leads to a nontrivial $f_+(s)$ in Eq.~\eqref{Eq:Mgen}, while in the ChPT analysis of Sec.~\ref{sec:ChPT} one has $f_+(s)=1$ at this order in the expansion.

\begin{figure}[t]
    \centering
              \includegraphics[width=0.4\linewidth]{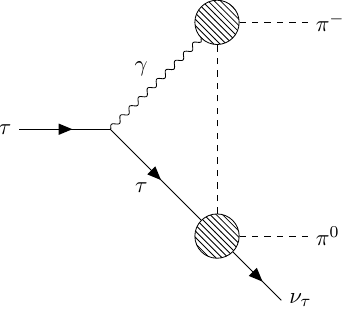}
    \caption{Triangle/box diagram with $\pi\pi\gamma$ and $\tau\nu_\tau\pi\pi$ vertices dressed with the pion VFF.  }
    \label{fig:disp_tri}
\end{figure}

For the calculation of radiative corrections, the most important modification beyond ChPT is represented in Fig.~\ref{fig:disp_tri}, showing the analog of Fig.~\ref{fig:loops}$(a)$, but with vertices dressed by the pion VFF. In a dispersive picture, such a contribution arises by considering the pion-pole singularity in the general matrix element $\langle \pi\pi^0|j_\text{em}^\mu j_W^\nu|0\rangle$ of weak and electromagnetic currents, which could be generalized in analogy to pion Compton scattering~\cite{Garcia-Martin:2010kyn,Hoferichter:2011wk,Moussallam:2013una,Danilkin:2018qfn,Hoferichter:2019nlq,Danilkin:2019opj}. However, at low energies the pion-pole contribution gives by far the dominant effect, leading to the picture in Fig.~\ref{fig:disp_tri}. Given the short-range weak vertex, the diagram appears as a triangle topology, but actually originates from a box diagram, as will become more explicit in the dispersive analysis. 

Following the strategy of Ref.~\cite{Colangelo:2022lzg}, we insert a dispersive representation for the pion VFF at each vertex. In principle, one would need to differentiate between $F_\pi^V(s)$ and $f_+(s)$ for electromagnetic and weak vertices, but since the difference is higher order in IB, we use 
\begin{equation}
    \label{Eq:FpiV_weakv}
    f_+(s) = \frac{1}{\pi} \int_{4 \mpi^2}^{\infty}\text{d}s'\, \frac{\Im f_+(s')}{s'-s}
\end{equation}
for both. The unsubtracted form~\eqref{Eq:FpiV_weakv} is chosen to ensure that UV divergences already cancel within the box diagram itself (in contrast to ChPT), at the expense of having to fulfill the sum rule
\begin{equation}
    \label{Eq:FpiV_weakv0}
    f_+(0) = \frac{1}{\pi} \int_{4 M_\pi^2}^{\infty}\text{d}s'\, \frac{\Im f_+(s')}{ s'} =1
\end{equation}
to respect charge conservation. Before turning to the calculation of the box diagram, we briefly discuss the dispersive representation of $f_+(s)$ that we will use in Sec.~\ref{sec:fits} for our fits to the $\tau\to\pi\pi\nu_\tau$ spectral function. 

\subsection[Pion vector form factor \texorpdfstring{$f_+(s)$}{}]{Pion vector form factor \texorpdfstring{$\boldsymbol{f_+(s)}$}{}}

For the fit to the $\tau$ spectral function, we use 
a dispersive representation similar to the one developed for $e^+e^-\to\pi^+\pi^-$ in Refs.~\cite{Colangelo:2018mtw,Colangelo:2020lcg,Colangelo:2022prz,Hoferichter:2023sli,Stoffer:2023gba,Leplumey:2025kvv}, but with explicit resonance contributions for $\rho'$, $\rho''$~\cite{Hoferichter:2014vra,Hoferichter:2019mqg,Hoferichter:2023bjm,Hoferichter:2025lcz}:
\beq
\label{eq:piFF}
f_+(s)=\bigg[1+G_\text{in}^N(s)+\sum_{V=\rho',\rho''}c_V{\mathcal A}_V(s)\bigg]\Omega^1_1(s)\,.
\eeq
The Omn\`es function~\cite{Omnes:1958hv} is defined as
\beq
\label{f+_disp}
\Omega^1_1(s)\equiv\exp\bigg\{\frac{s}{\pi}\int_{4\mpi^2}^\infty \text{d}s'\,\frac{\delta_1^1(s')}{s'(s'-s)}\bigg\}\,,
\eeq
where the $P$-wave $\pi\pi$ phase shift $\delta^1_1(s)$ is parameterized as the solution of the Roy equations~\cite{Roy:1971tc,Ananthanarayan:2000ht,Caprini:2011ky} in terms of the phase-shift values at $s_0=(0.8\GeV)^2$, $s_1=(1.15\GeV)^2$. Above $\sqrt{s_c^\delta}=1.3\GeV$, the phase shift is continued to the value of $\pi$ with the prescription
\begin{equation}
    \delta(s)=\pi-\frac{\left(\pi-\delta(s_c^\delta)\right)^2}{\pi-\delta(s_c^\delta)+\delta'(s_c^\delta)(s-s_c^\delta)}\, ,
\end{equation}
ensuring that $\delta(s)$ is continuously differentiable at $s_c^\delta$.

Inelastic contributions are expanded into a conformal polynomial
\begin{equation}
    G_\text{in}^N(s)=\sum_{k=1}^Np_k\left(\big[z(s)\big]^k-\big[z(0)\big]^k\right) \, ,
\end{equation}
with conformal variable
\begin{equation}
    z(s)=\frac{\sqrt{s_\text{in}-s_c}-\sqrt{s_\text{in}-s}}{\sqrt{s_\text{in}-s_c}+\sqrt{s_\text{in}-s}}\, ,
\end{equation}
which should account for inelastic channels such as $4\pi$.
In view of the phenomenological finding that inelasticities below the $\pi\omega$ threshold are negligible, we set $s_\text{in}=(M_\omega+M_\pi)^2$.\footnote{For $s>s_\text{in}$, the square root in $z(s)$ can be analytically continued as $\sqrt{s_\text{in}-s}\to-i\sqrt{s-s_\text{in}}$, in such a way that $z(s)$ matches the behavior of $z(s+i\epsilon)$ with infinitesimally small $\epsilon>0$.} Furthermore, the correct threshold behavior of $G_\text{in}^N(s)$ is enforced by removing the term $\propto\sqrt{s_\text{in}-s}$ by means of the condition~\cite{Colangelo:2018mtw}
\begin{equation}
    p_1=-\sum_{k=2}^Nk\,p_k\, ,
\end{equation}
leaving $N-1$ free parameters in $G_\text{in}^N(s)$. The point that is mapped to the origin in the conformal map is identified as $s_c$, which is set to $s_c=-1\,\text{GeV}^2$ in the following.

The last term in Eq.~\eqref{f+_disp} takes into account the contribution of the $\rho'\equiv\rho(1450)$ and $\rho''\equiv\rho(1700)$ resonances in the $\pi\pi$ spectral function. It is given by
\begin{align}
\label{Arhoprhopp}
{\mathcal A}_V(s)&=\frac{s}{\pi}\int_{s_\text{in}}^\infty \text{d}s'\,\frac{\Im {\mathcal A}_V(s')}{s'(s'-s)}\,,\notag\\
\Im {\mathcal A}_V(s)&=\Im\frac{1}{M_V^2-s-i\sqrt{s} \Gamma_V(s)}\,,
\end{align}
where the energy-dependent widths are constructed from the $\omega\pi$ phase space~\cite{Zanke:2021wiq}
\begin{equation}
    \hat{\Gamma}_V(s)=\Gamma_V\frac{\gamma_{V\to\omega\pi}(s)}{\gamma_{V\to\omega\pi}(M_V^2)}\theta\left(s-s_\text{in}\right)\, ,\quad\gamma_{V\to\omega\pi}(s)=\frac{\lambda^{3/2}(s,M_\omega^2,M_\pi^2)}{s^{3/2}}\,,
\end{equation}
neglecting the $V\to\pi\pi$ and $V\to a_1\pi\;(a_1\to3\pi)$ decay channels of $\rho'$ and $\rho''$. In addition, centrifugal barrier factors according to \cite{COMPASS:2015gxz,VonHippel:1972fg}
\begin{equation}
    \Gamma_V(s)=\hat{\Gamma}_V(s)\frac{s}{M_V^2}\frac{\lambda(M_V^2,M_\omega^2,M_\pi^2)+4M_V^2p_R^2}{\lambda(s,M_\omega^2,M_\pi^2)+4s\,p_R^2}\, ,\quad p_R=202.4\,\text{MeV}\, ,
\end{equation}
are implemented, to tame the asymptotic behavior of the energy-dependent widths. We also studied representations in which $\rho'$, $\rho''$ are introduced in the conformal variable~\cite{Kirk:2024oyl}, but found that the proximity of the $\rho''$ to the edge of the phase space makes it very hard to find stable fits, even though eventually a parameterization in terms of the pole parameters instead of Breit--Wigner parameters as in Eq.~\eqref{Arhoprhopp} would be preferable.

Because of the use of unsubtracted dispersion relation in the calculation of virtual corrections, the sum rule in Eq.~\eqref{Eq:FpiV_weakv0} needs to be imposed as additional constraint, which we incorporate again via the coefficient of $G_\text{in}^N$(s). Explicitly, 
setting 
\begin{equation}
    p_2=\frac{1+\frac{1}{\pi}\int_{4M_\pi^2}^\infty\text{d}s\,C_N(s)}{\frac{1}{\pi}\int_{4M_\pi^2}^\infty\text{d}s\,C_D(s)}\,,
\end{equation}
with
\begin{align}
    C_N(s)&=\frac{1}{s}\Im\Bigg\{\Omega^1_1(s)\bigg[\sum_{k=3}^Nk\,p_k\big(z(s)-z(0)\big)-\sum_{k=3}^Np_k\left(\big[z(s)\big]^k-\big[z(0)\big]^k\right)\notag\\
    &\qquad-\sum_{V=\rho',\rho''}c_V\mathcal{A}_{V}(s)-1\bigg]\Bigg\}\, ,\notag\\
    C_D(s)&=\frac{1}{s}\Im\left\{\Omega^1_1(s)\Big[\big[z(s)\big]^2-\big[z(0)\big]^2-2\big(z(s)-z(0)\big)\Big]\right\}\, ,
\end{align}
the resulting $f_+(s)$ satisfies the normalization condition~\eqref{Eq:FpiV_weakv0}. 

\subsection{Dispersive calculation of the box diagram}
\label{sec:BoxFF}

When supplementing the box diagram with unsubtracted dispersive representations of the pion VFF, see Eq.~\eqref{Eq:FpiV_weakv}, and projecting it onto the $\tilde f_+(s,t)$ form factor, see Eq.~\eqref{Eq:Mgen}, the following expression can be obtained
\begin{align}
\label{Eq:fplus_pavedecomp}
    \tilde f_+^{\text{disp}}(s,t) = \frac{e^2}{16\pi^4f_+(s)}\int_{4M_\pi^2}^\infty \text{d}s'\int_{4\mpi^2}^\infty \text{d}s''\, \big[ \tilde{f}_B(s,s',s'') + \tilde{f}_C(s,t,s',s'') + \tilde{f}_D(s,t,s',s'') \big]\, ,
\end{align}
where
\begin{align}
\label{eq:fBCD}
    \tilde{f}_B &= \frac{\Im f_+(s') \Im f_+(s'')}{2 s' (s - 4M_\pi^2)} \big[ \bar{B}_0(M_\pi^2,M_\pi^2,s') -2 \bar{B}_0(s,s',s'')\big] \, ,\notag\\
    \tilde{f}_C&= \frac{\Im f_+(s') \Im f_+(s'')}{2 s' \big[s ((t-M_\pi^2)^2 + st) + M_\pi^2 m_\tau^4 - m_\tau^2 s (M_\pi^2 + t) \big]} \notag \\
    &\times \bigg\{ -\big[2 s t (t-M_\pi^2) + m_\tau^6 + m_\tau^4 (2M_\pi^2 - s -2t) - m_\tau^2(M_\pi^2-t)(M_\pi^2-s+t)\big]\notag\\
    &\qquad\times\bar{C}_0(m_\tau^2,M_\pi^2,t,m_\tau^2,s',M_\pi^2) \notag \\
    &+\big[2 s (2 M_\pi^4 - 4M_\pi^2 t + t(s+2t)) + m_\tau^6 + m_\tau^4 (3M_\pi^2 - 2s - t)+ m_\tau^2 s (s - t - 3M_\pi^2) \big]\notag\\
    &\qquad\times \bar{C}_0(0,m_\tau^2, s, s'', m_\tau^2,s') \notag\\
    &-\Big[s \big\{M_\pi^4 (s-s''-16 t)+2 M_\pi^2 t (5 s+s''+8 t)-t (t (3 s+s'')+s (s+s''))\big\}\notag\\
    &\qquad+m_\tau^4 \left(8 M_\pi^4+M_\pi^2 (3 s-s'')-s^2\right)\notag\\
    &\qquad+m_\tau^2 s \big(4 M_\pi^4+M_\pi^2 (s''-6 s-20 t) +s^2+4 s t+s'' t\big)\Big]\frac{\bar{C}_0(M_\pi^2,M_\pi^2,s,s',M_\pi^2,s'')}{s-4M_\pi^2}\notag \\
    &\quad- \frac{s'\big(m_\tau^2 s (M_\pi^2 +t) - m_\tau^4M_\pi^2 - s((t-M_\pi^2)^2 + st) \big)}{s-4M_\pi^2} C_0(M_\pi^2,M_\pi^2,s,s',M_\pi^2,s'') \bigg\}\,,\notag\\
    \tilde{f}_D&= \frac{\Im f_+(s') \Im f_+(s'')}{2 s' \big[s((t-M_\pi^2)^2 + st) + M_\pi^2 m_\tau^4 - m_\tau^2 s (M_\pi^2+t) \big]}\notag\\
    &\times\bigg\{ s'(t-M_\pi^2)\big[m_\tau^2(M_\pi^2+s+t)-2st-m_\tau^4\big]D_0(m_\tau^2,M_\pi^2,M_\pi^2,0,t,s,m_\tau^2,s',M_\pi^2,s'')\notag\\
   &+\Big[m_\tau^6 \left(4 M_\pi^2+s-s''\right) +m_\tau^4 \left(4 M_\pi^4-2 M_\pi^2 (s+s''+2 t)-s^2+s s''-6 s t+2 s'' t\right)\notag\\
    &\qquad +m_\tau^2 \big(M_\pi^4 (s''-s)+M_\pi^2 s (s-s''-8 t)+t \big(3 s^2+s s''+9 s t-s'' t\big)\big)\notag\\
    &\qquad +2 s \left(M_\pi^2-t\right) \left(2 M_\pi^4-4 M_\pi^2 t+t (s+s''+2 t)\right)\Big]\notag\\
    &\qquad\times\bar{D}_0(m_\tau^2,M_\pi^2,M_\pi^2,0,t,s,m_\tau^2,s',M_\pi^2,s'')\bigg\} \, ,
\end{align}
and we suppressed the arguments of $\tilde{f}_B(s,s',s'')$ etc.\ for brevity. The $B_0$, $C_0$, $D_0$ loop functions are given in the FeynCalc convention, with subtracted variants defined by
\begin{equation}
\label{BCD_sub}
    \bar{F}_0(\ldots,s',\ldots)\equiv F_0(\ldots,s',\ldots)-F_0(\ldots,0,\ldots)\, ,\qquad F\in\{B,C,D\}\,.
\end{equation}
In contrast to the ChPT analysis, the form factor $\tilde f_+(s,t)$ now indeed depends on both Mandelstam variables.  Moreover, thanks to the use of the unsubtracted dispersive representation of $f_+(s)$, the result is now manifestly UV finite, i.e., the suppression of the high-energy modes due to the additional propagators suffices to render the loop integrals finite. 
Finally, an important observation concerns 
the common factor appearing in the denominators of $\tilde{f}_C$ and $\tilde{f}_D$, which can be written as
\begin{equation}
    s((t-M_\pi^2)^2 + st) + M_\pi^2 m_\tau^4 - m_\tau^2 s (M_\pi^2+t) = s(t - t_\text{min})(t - t_\text{max}) \, ,\label{Eq:endpointsing}
\end{equation}
for $t_\text{min/max}$ of Eq.~\eqref{tminmax} in the isospin limit. These divergences at the border of the phase space are canceled by corresponding zeros in the numerator, but the cancellation needs to be made explicit for a stable implementation, see Sec.~\ref{sec:endpoint}. 

\subsection[Numerical treatment of \texorpdfstring{$D_0$}{} functions]{Numerical treatment of \texorpdfstring{$\boldsymbol{D_0}$}{} functions}
\label{sec:D0}

The scalar integral $D_0(m_\tau^2,M_\pi^2,M_\pi^2,0,t,s,m_\tau^2,0,M_\pi^2,s'')$, appearing in the expression for the $\tilde f_+^\text{disp}(s,t)$ form factor, exhibits a singularity at $s'' = s$, introducing an obstacle for the  numerical evaluation of the dispersion integrals in Eq.~\eqref{Eq:fplus_pavedecomp}. In the scalar integral, this divergent part appears in 
\begin{equation}
D_0(m_\tau^2,M_\pi^2,M_\pi^2,0,t,s,m_\tau^2,0,M_\pi^2,s'')=-\frac{d_0(t)}{s''-s}\bigg(32\pi^2\Lambda_\text{IR}+1-\log\frac{\mu_\text{IR}^2}{\mpi m_\tau}\bigg)    
    +D_0^{\pi\tau}(s,t,s'') \, ,
\end{equation}
where~\cite{Beenakker:1988jr,Ellis:2007qk}
\begin{align}
 D_0^{\pi\tau}(s,t,s'') &\equiv \texttt{ScalarD0IR16}[0,\mpi^2,s,t,m_\tau,\sqrt{s''},\mpi]\notag\\
 &=\frac{2d_0(t)}{s''-s}\log\frac{s''}{s''-s}
+D_0^\text{rest}(t,s'')\,,\notag\\
 D_0^\text{rest}(t,s'')&=\frac{1}{s''-s}\frac{1}{\lambda_{\pi\tau}^{1/2}(t)}\bigg[-\log x_t\log\frac{\mpi m_\tau}{s''}+\log^2 x_1+\log^2 x_2\notag\\
 &-\text{Li}_2\big(1-x_t^2\big)+\sum_{y\in\big\{x_1x_2,\frac{1}{x_1x_2},\frac{x_1}{x_2},\frac{x_2}{x_1}\big\}}\text{Li}_2\big(1-x_t y\big)\bigg]\,,\notag\\
 d_0(t) &=-\frac{t}{\lambda_{\pi\tau}(t)}B_0^{\pi\tau}(t) \, ,
\qquad x_1=\frac{\sqrt{s''}}{m_\tau}\,,\qquad 
x_2=\bigg(\frac{1-\sigma_\pi(s'')}{1+\sigma_\pi(s'')}\bigg)^{1/2}\,.
\end{align}
The dispersion integral over $s''$ can then be written in the following way
\begin{align}
    \tilde f_+^\text{disp}(s,t) \supset \int_{4M_\pi^2}^{\infty} \text{d}s'' \, \Im f_+(s'') \frac{p_1(s,t) + p_2(s,t) s''}{s'' - s}\left[2 d_0(t) \log \frac{s^{\prime \prime }}{s'' - s} + D_0^\text{rest}(t,s'') \right] \, ,
\end{align}
where $p_1$ and $p_2$ collect the kinematic prefactors independent of $s''$. In order to facilitate the numerical integration, we add and subtract $\Im f_+(s)$ multiplied with the divergent part of the integrand. By doing so, we obtain
\begin{align}
\label{Eq:D0spectreat}
    \tilde f_+^\text{disp}(s,t) &\supset \int_{4M_\pi^2}^{\infty} \mathrm{d}s'' \, \big\{ \Im f_+(s'') - \Im f_+(s) \big\}\frac{p_1(s,t) + p_2(s,t) s}{s'' - s} \\
    &\qquad\times\left[2 d_0(t) \log \frac{s^{\prime \prime}}{s'' - s} + D_0^\text{rest}(t,s'') \right] \notag \\
    &+p_2(s,t)\int_{4M_\pi^2}^{\infty} \mathrm{d}s'' \, \Big\{ \Im f_+(s'') - \Im f_+(s) \theta(\Lambda^2-s'')\Big\}\notag\\
    &\qquad\times\left[2 d_0(t) \log \frac{s^{\prime \prime}}{s'' - s} + D_0^\text{rest}(t,s'') \right]\notag\\
    &+ \Im f_+(s) \Bigg\lbrace d_0(t) \bigg[p_2(s,t) I_{\ell 1}(s,\Lambda^2) + \big(p_1(s,t)+p_2(s,t) s\big) I_{\ell2}(s) \bigg] \notag\\
    &\qquad+\big(p_1(s,t)+p_2(s,t) s\big) \int_{4M_\pi^2}^{\infty} \mathrm{d}s'' \, \frac{D_0^\text{rest}(t,s'')}{s''-s}+ p_2(s,t) \int_{4M_\pi^2}^{\Lambda^2} \text{d}s'' \, D_0^\text{rest}(t,s'')\Bigg\rbrace\, , \notag
\end{align}
written in terms of the analytically calculated integrals
\begin{align}
    I_{\ell 1}(s,\Lambda^2) &= 2\int_{4M_\pi^2}^{\Lambda^2} \mathrm{d}s''\, \log \frac{s^{\prime \prime}}{s'' - s} \notag \\
    &=2 \bigg\lbrace s \log \frac{\Lambda^2 - s}{s - 4M_\pi^2}+ \Lambda^2 \log \frac{\Lambda^2}{\Lambda^2 - s}+ 4 M_\pi^2 \log \frac{s - 4M_\pi^2}{4M_\pi^2} + i \pi (s - 4M_\pi^2) \bigg\rbrace\, , \notag\\
    I_{\ell2}(s) &= 2\int_{4M_\pi^2}^{\infty} \mathrm{d}s''\, \frac{1}{s'' - s - i \epsilon} \log \frac{s^{\prime \prime}}{s'' - s - i \epsilon}\notag\\
    &= 2 \text{Li}_2\left( 1 - \frac{4M_\pi^2}{s} \right)  - \frac{2 \pi^2}{3} + \left(2 \pi i + \log \frac{s}{s-4M_\pi^2} \right) \log \frac{s}{s - 4M_\pi^2} \, .
    \label{D0_integrals_analytic}
\end{align}
The dependence on the (high-energy) integral cutoff $\Lambda^2$ of $I_{\ell 1}$ and the one in the integral $\int \text{d}s'' D_0^\text{rest}(t,s'')$ in Eq.~\eqref{Eq:D0spectreat} cancels against the $\Lambda$ dependence introduced via $\theta(\Lambda^2-s'')$ in the second term proportional to $p_2(s,t)$. The imaginary part that emerges from the Cauchy propagator, see Eq.~\eqref{D0_integrals_analytic}, also needs to be kept, as it contributes to $\Re\tilde f_+(s,t)$ due to the imaginary part of $f_+(s)$ in Eq.~\eqref{Eq:fplus_pavedecomp}.

\subsection{Endpoint singularities in the phase space}
\label{sec:endpoint}

As anticipated in Sec.~\ref{sec:BoxFF}, the common denominator in the expression of the box diagram~\eqref{Eq:endpointsing} yields an endpoint singularity once the phase-space integral in $t$ is performed. To show that this divergence is canceled by the numerator and derive a numerically stable formulation, we proceed as follows. As first step we split the box form factor $ \tilde f_+^{\text{disp}}(s,t)$  as given in Eq.~\eqref{Eq:fplus_pavedecomp} into two contributions
\begin{equation}
     \tilde f_+^{\text{disp}}(s,t) =
     \frac{e^2}{16\pi^4f_+(s)}\int_{4M_\pi^2}^\infty \text{d}s'\int_{4\mpi^2}^\infty \text{d}s''\,\bigg[\tilde f_+^\text{fin}(s,t,s',s'')+\frac{N(s,t,s',s'')}{s(t-t_{\text{min}})(t-t_{\text{max}})}\bigg]\, ,\label{eq:f+_rearr}
\end{equation}
where $\tilde f_+^\text{fin}(s,t,s',s'')$ includes all the terms of $\tilde f_+^{\text{disp}}(s,t)$ without the divergent denominator and corresponds to $\tilde{f}_B(s,s',s'')$ in Eq.~\eqref{eq:fBCD}. In order to show that the divergences at $t-t_{\text{min/max}}$ explicitly cancel, we aim to rearrange the numerator as $N(s,t,s',s'')=(t-t_{\text{min}})(t-t_{\text{max}})\bar{N}(s,t,s',s'')$.
First, we observe that
$N(s,t_{\text{min}},s',s'')=N(s,t_{\text{max}},s',s'')=0$, which can be proved analytically using the algorithm provided in Ref.~\cite{Denner:1991kt} to factorize $D_0$ at the phase-space boundary into simpler loop functions
\begin{align}
&D_0(m_\tau^2,M_\pi^2,M_\pi^2,0,t,s,m_\tau^2,s',M_\pi^2,s'')\big|_{t=t_{\text{min}/\text{max}}}\notag\\
&=\frac{1}{\mathcal{D}(s,t,s',s'')}\Bigg\{\bigg[m_\tau^4\left(s''-s\right)+s\left(s-s'-s''\right)\left(t-\mpi^2\right)\notag\\
   &\qquad\qquad+m_\tau^2\Big(s(s-s'')+(s+s'-s'')(t-\mpi^2)\Big)\bigg]C_0(0,m_\tau^2,s,s'',m_\tau^2,s')\notag\\
   &\qquad+\Big[2\big(m_\tau^4\mpi^2+ss''t\big)+(s-s'+s'')(\mpi^2-t)^2-m_\tau^2(s+s'')(t+\mpi^2)\Big]\notag\\
   &\qquad\qquad\times C_0(0,M_\pi^2,t,m_\tau^2,s'',M_\pi^2)\notag\\
   &\qquad+\Big[m_\tau^4(s-s'')+(s+s'-s'')(\mpi^2-t)^2+2ss't-m_\tau^2(2s+s'-2s'')(t+\mpi^2)\Big]\notag\\
   &\qquad\qquad\times C_0(m_\tau^2,M_\pi^2,t,m_\tau^2,s',M_\pi^2)\notag\\
   &\qquad+\Big[m_\tau^2\Big(2\mpi^2(s+s'-s'')+s(s''-s)\Big)+s(s-s'-s'')(t+\mpi^2)\Big]\notag\\
   &\qquad\qquad\times C_0(M_\pi^2,M_\pi^2,s,s',M_\pi^2,s'')\Bigg\} \Bigg|_{t=t_{\text{min}/\text{max}}} \,,
\end{align}
with
\begin{align}
    \mathcal{D}(s,t,s',s'')&=
  \lambda_{\pi\tau}(t)\lambda(s,s',s'')\notag\\  &+s'\Big[m_\tau^4\left(s'-4\mpi^2\right)+2m_\tau^2\left(s-s'+s''\right)\left(m_\tau^2-\mpi^2-t\right)-4s s''t\Big]\,,
\end{align}
and likewise for the second $D_0$ function with $s'=0$. Inserting these relations into Eqs.~\eqref{Eq:fplus_pavedecomp}
and~\eqref{eq:fBCD}, indeed the numerators vanish, canceling the 
zero in the denominator~\eqref{Eq:endpointsing}. 
Accordingly, we can write
\begin{equation}
    N(s,t,s',s'')=N(s,t,s',s'')-N(s,t_{\text{max}},s',s'')\equiv(t-t_{\text{max}})N_+(s,t,s',s'')\, ,
\end{equation}
which defines $N_+(s,t,s',s'')$. Then
\begin{align}\label{eq:num}
    N(s,t,s',s'')&=(t-t_{\text{max}})N_+(s,t,s',s'')\notag\\
    &=(t-t_{\text{max}})\big[N_+(s,t,s',s'')-N_+(s,t_{\text{min}},s',s'')\big]\notag\\
    &\equiv(t-t_{\text{max}})(t-t_{\text{min}})\bar{N}(s,t,s',s'') \, .
\end{align}
After this rearrangement, the $\bar{N}(s,t,s',s'')$ in Eq.~\eqref{eq:num} can be divided into three contributions: $\bar{N}(s,t,s',s'')=\bar{N}^\text{fin}(s,t,s',s'')+\bar{N}_+(s,t,s',s'')+\bar{N}_-(s,t,s',s'')$. The first term does not involve the endpoint singularities anymore and can be grouped in $\tilde f_+^\text{fin}(s,t,s',s'')$, so that
\begin{align}
     \tilde f_+^\text{fin}&=\frac{\Im f_+(s')\Im f_+(s'')}{s' }\Bigg\{\frac{\bar{B}_0(M_\pi^2,M_\pi^2,s')-2\bar{B}_0(s,s',s'')}{2(s-4M_\pi^2)}+2\bar{C}_0(0,m_\tau^2,s,s'',m_\tau^2,s')\notag\\
     &+\frac{C_0(M_\pi^2,M_\pi^2,s,s',M_\pi^2,s'')}{2(s-4M_\pi^2)}+\frac{3s+s''-16M_\pi^2}{2(s-4M_\pi^2)}\bar{C}_0(M_\pi^2,M_\pi^2,s,s',M_\pi^2,s'')\notag\\
     &-\frac{2s+m_\tau^2}{2s}\bar{C}_0(m_\tau^2,M_\pi^2,t,m_\tau^2,s',M_\pi^2)\notag\\
     &+\frac{m_\tau^2-2s}{2s}D_0(m_\tau^2,M_\pi^2,M_\pi^2,0,t,s,m_\tau^2,s',M_\pi^2,s'')\notag\\
     &-\frac{m_\tau^2(s''-9s)+2s\left[s+s''-6M_\pi^2+2(t+t_{\text{min}}+t_{\text{max}})\right]}{2s}\notag\\
     &\qquad \times\bar{D}_0(m_\tau^2,M_\pi^2,M_\pi^2,0,t,s,m_\tau^2,s',M_\pi^2,s'')\Bigg\}\, .
\end{align}
The other two terms are
\begin{align}
    \bar{N}_+&=\frac{\Im f_+(s')\Im f_+(s'')}{(t-t_{\text{max}})(t_{\text{max}}-t_{\text{min}}) }\Bigg\{\frac{\bar{C}_0^+(m_\tau^2,M_\pi^2,t,m_\tau^2,s',M_\pi^2)}{2s s'}\bigg[2s t_{\text{max}}(M_\pi^2-t_{\text{max}})\notag\\
    &\qquad+m_\tau^2(M_\pi^2-t_{\text{max}})(M_\pi^2-s+t_{\text{max}}) +m_\tau^4(s+2t_{\text{max}}-2M_\pi^2)-m_\tau^6\bigg]\notag\\
    &-\frac{(M_\pi^2-t_{\text{max}})\left[m_\tau^2(M_\pi^2+s+t_{\text{max}})-m_\tau^4-2s t_{\text{max}}\right]}{2s} \notag\\
    &\qquad\times D_0^+(m_\tau^2,M_\pi^2,M_\pi^2,0,t,s,m_\tau^2,s',M_\pi^2,s'')\notag\\
    &+\frac{\bar{D}_0^+(m_\tau^2,M_\pi^2,M_\pi^2,0,t,s,m_\tau^2,s',M_\pi^2,s'')}{2s s'}\bigg[m_\tau^6(4M_\pi^2+s-s'')\notag\\
    &\qquad +m_\tau^4\left(4M_\pi^2-2M_\pi^2(s+s''+2t_{\text{max}})-s^2+s s''-6s t_{\text{max}}+2s'' t_{\text{max}}\right)\notag\\
    &\qquad+m_\tau^2\left(M_\pi^4(s''-s)+M_\pi^2 s(s-s''-8t_{\text{max}})+t_{\text{max}}(3s^2+s s''+(9s-s'') t_{\text{max}})\right)\notag\\
    &\qquad +2s(M_\pi^2-t_{\text{max}})\left(2M_\pi^4-4M_\pi^2t_{\text{max}}+t_{\text{max}}(s+s''+2t_{\text{max}})\right)\bigg]\Bigg\}\, ,
\end{align}
and
\begin{align}
    \bar{N}_-&=\frac{\Im f_+(s')\Im f_+(s'')}{(t-t_{\text{min}})(t_{\text{min}}-t_{\text{max}}) }\Bigg\{\frac{\bar{C}_0^-(m_\tau^2,M_\pi^2,t,m_\tau^2,s',M_\pi^2)}{2s s'}\bigg[2s t_{\text{min}}(M_\pi^2-t_{\text{min}})\notag\\
    &\qquad+m_\tau^2(M_\pi^2-t_{\text{min}})(M_\pi^2-s+t_{\text{min}}) +m_\tau^4(s+2t_{\text{min}}-2M_\pi^2)-m_\tau^6\bigg]\notag\\
    &-\frac{(M_\pi^2-t_{\text{min}})\left[m_\tau^2(M_\pi^2+s+t_{\text{min}})-m_\tau^4-2s t_{\text{min}}\right]}{2s} \notag\\
    &\qquad\times D_0^-(m_\tau^2,M_\pi^2,M_\pi^2,0,t,s,m_\tau^2,s',M_\pi^2,s'')\notag\\
    &+\frac{\bar{D}_0^-(m_\tau^2,M_\pi^2,M_\pi^2,0,t,s,m_\tau^2,s',M_\pi^2,s'')}{2s s'}\bigg[m_\tau^6(4M_\pi^2+s-s'')\notag\\
    &\qquad +m_\tau^4\left(4M_\pi^2-2M_\pi^2(s+s''+2t_{\text{min}})-s^2+s s''-6s t_{\text{min}}+2s'' t_{\text{min}}\right)\notag\\
    &\qquad+m_\tau^2\left(M_\pi^4(s''-s)+M_\pi^2 s(s-s''-8t_{\text{min}})+t_{\text{min}}(3s^2+s s''+(9s-s'') t_{\text{min}})\right)\notag\\
    &\qquad+2s(M_\pi^2-t_{\text{min}})\left(2M_\pi^4-4M_\pi^2t_{\text{min}}+t_{\text{min}}(s+s''+2t_{\text{min}})\right)\bigg]\Bigg\}\, ,
\end{align}
where
\begin{equation}
    F_0^\pm(\ldots,t,\ldots)\equiv F_0(\ldots,t,\ldots)-F_0(\ldots,t_{\text{max/min}},\ldots)\,,
\end{equation}
and we again suppressed the arguments of $\tilde f_+^\text{fin}(s,t,s',s'')$ and $\bar N_\pm(s,t,s',s'')$ for brevity. 
In this form, the latter still have a divergent denominator at $t_{\text{max/min}}$, respectively, but these remaining (removable) singularities can be dealt with by 
 expanding  around $t=t_{\text{max/min}}$ when the integration in $t$ is close to the boundaries $t_{\text{max/min}}$.

\subsection{IR singularities and low-energy limit}
\label{sec:IR}

The $D_0$ function discussed in Sec.~\ref{sec:D0} gives an example of an IR-divergent loop function in Eq.~\eqref{eq:fBCD}. In general, it is the subtraction of the $s'=0$ terms in Eq.~\eqref{BCD_sub} that can introduce IR divergences, a second one arises in 
\begin{equation}
    C_0(m_\tau^2,M_\pi^2,t,m_\tau^2,0,M_\pi^2)=C_0^{\pi\tau}(t)-\frac{t B_0^{\pi\tau}(t)}{\lambda_{\pi\tau}(t)}\Big(32\pi^2\Lambda_\text{IR}+1-\log\frac{\mu_\text{IR}^2}{\mpi m_\tau}\Big)\, ,
\end{equation}
which is the same loop function that defines the bulk of ChPT result~\eqref{eq:fpBoxChpt}. 
For both the $D_0$ and $C_0$ terms, the only remaining dependence on $s'$ disappears via the sum rule~\eqref{Eq:FpiV_weakv0}, and, upon separating $s''=(s''-s)+s$, the $D_0$ terms in which $s''$ appears in the numerator cancel exactly the $C_0$ contribution, leaving 
\begin{align}
    \tilde f^\text{disp}_+(s,t)\Big|_\text{IR}&=-\frac{e^2}{16\pi^2 f_+(s)}\frac{1}{\pi}\int_{4\mpi^2}^\infty \text{d}s''\, \frac{\Im f_+(s'')}{s''-s} \frac{2t (\mpi^2+m_\tau^2-t)B_0^{\pi\tau}(t)}{\lambda_{\pi\tau}(t)}32\pi^2\Lambda_\text{IR}\notag\\
    &=-\frac{e^2}{16\pi^2} \frac{2t (\mpi^2+m_\tau^2-t)B_0^{\pi\tau}(t)}{\lambda_{\pi\tau}(t)}32\pi^2\Lambda_\text{IR}\,,
\end{align}
after inserting Eq.~\eqref{Eq:FpiV_weakv}. The resulting IR divergence therefore matches exactly the IR singularity of $\tilde f_+^{(a)}(s,t)$ obtained in  ChPT, see Eq.~\eqref{eq:fpBoxChpt}. 

This observation is in fact more general: it is always possible to apply the chiral counting to a dispersive representation, for example by expanding the integrands at low energy. By construction, this reproduces the ChPT representation order by order, as long as the same set of pionic intermediate states is included.
One key aspect concerns the IR structure, another one the chiral logarithms. To illustrate the latter point, we consider the expansion around $s=t=0$
\begin{align}
\label{f0a_ChPT}
\tilde f_+^{(a)}(0,0)&=\frac{e^2}{16\pi^2}
    \bigg[\frac{m_\tau^2-2\mpi^2}{m_\tau^2-\mpi^2}+\frac{\mpi^4+2\mpi^2m_\tau^2-7m_\tau^4}{4(m_\tau^2-\mpi^2)^2}\log\frac{\mpi^2}{m_\tau^2}-\frac{7}{4}\Big(32\pi^2\Lambda_\text{UV}-\log\frac{\mu_\text{UV}^2}{m_\tau^2}\Big)\notag\\
&+\frac{m_\tau^2+\mpi^2}{m_\tau^2-\mpi^2}\log\frac{\mpi^2}{m_\tau^2}\Big(32\pi^2\Lambda_\text{IR}+1-\log\frac{\mu_\text{IR}^2}{\mpi m_\tau}\Big)
\bigg]\,,    
\end{align}
where we have written the UV scale with respect to $m_\tau$ to isolate the chiral logarithm. Using a narrow-width approximation $\Im f_+(s)=\pi\delta(s-M_\rho^2)M_\rho^2$, the dispersive result for $s=t=0$ can be performed in terms of polylogarithms, and the expansion for $M_\rho\to\infty$ gives 
\begin{align}
\label{f0disp}
\tilde f_+^\text{disp}(0,0)&=\frac{e^2}{16\pi^2}
    \bigg[\frac{3}{8}-\frac{m_\tau^2}{m_\tau^2-\mpi^2}+\frac{\mpi^4+2\mpi^2m_\tau^2-7m_\tau^4}{4(m_\tau^2-\mpi^2)^2}\log\frac{\mpi^2}{m_\tau^2}+\frac{7}{4}\log\frac{M_\rho^2}{m_\tau^2}\notag\\
&+\frac{m_\tau^2+\mpi^2}{m_\tau^2-\mpi^2}\log\frac{\mpi^2}{m_\tau^2}\Big(32\pi^2\Lambda_\text{IR}+1-\log\frac{\mu_\text{IR}^2}{\mpi m_\tau}\Big)
\bigg]+\Order\Big(M_\rho^{-2}\Big)\,. 
\end{align}
The comparison of Eqs.~\eqref{f0a_ChPT} and~\eqref{f0disp} shows again that the IR singularities coincide, but also that the coefficients of the chiral logarithm match exactly. Moreover, the coefficient of the UV divergence is reproduced, upon identifying $\mu_\text{UV}=M_\rho$, so that, as expected, the result again becomes UV divergent in the pointlike limit $M_\rho\to\infty$. The only difference then concerns finite contact terms, related to LECs in ChPT.  

Using the narrow-width limit, one can also study the threshold behavior of the resulting $G_\text{EM}(s)$ function analytically. In particular, we showed that the same factorization of $D_0$ functions based on which the endpoint singularities in Sec.~\ref{sec:endpoint} disappear, also ensures that $G_\text{EM}(s)$ derived from the IR-finite parts of $\tilde f_+^\text{disp}(s,t)$ remains finite at threshold. Accordingly, singularities at threshold only originate from the remainder of the IR cancellation, see Sec.~\ref{sec:threshold_gLow}, which behaves as $\log(s-4\mpi^2)$, and real-emission diagrams, the latter leading to the dominant $1/(s-4\mpi^2)$ threshold divergence.

\section{Matching}
\label{sec:match}

As discussed in Sec.~\ref{sec:IR}, our dispersive representation for the box diagram, Fig.~\ref{fig:disp_tri}, reproduces the IR properties of the ChPT diagram Fig.~\ref{fig:loops}$(a)$, i.e., the IR singularities and chiral logarithms. At the same time, the suppression due to the form factors in the loop integral ensures that the result is already manifestly UV finite. For a complete calculation, however, this improved representation for the box diagram needs to be combined with the full ChPT calculation 
\beq
\tilde f_+^\text{ChPT}(s,t)=\tilde f_+^{(a)}(s,t)+\tilde f_+^{(b)}(s,t)+\tilde f_+^{(c)}(s,t)+\tilde f_+^{\text{CT}+\text{SE}}(s,t)\,,
\eeq
see Eqs.~\eqref{eq:fpBoxChpt} and \eqref{CT+SE}. From the perspective of a dispersive analysis, all terms apart from $\tilde f_+^{(a)}(s,t)$ amount to a subtraction constant, as these diagrams do not involve a nontrivial dependence on the Mandelstam variables.  
Accordingly, we define our final result as
 \begin{equation}\label{eq:f_match}
        \tilde f_+^{\text{match}}(s,t)=\tilde f_+^{\text{disp}}(s,t)-\tilde f_+^{\text{disp}}(0,0)+\tilde f_+^{\text{ChPT}}(0,0)  \, .
    \end{equation}
With this procedure the final form factor $\tilde f_+^{\text{match}}(s,t)$ is still UV finite, has the right IR singularities, chiral logarithms, and, as an added benefit, displays a reduced sensitivity to the high-energy part of the dispersive integrals thanks to the subtraction at $s=t=0$.  
The explicit form of the subtraction term reads 
\begin{align}
    \tilde f_+^{\text{ChPT}}(0,0)&=\frac{e^2}{16\pi^2}\Bigg\{
    \frac{m_\tau^2-2\mpi^2}{m_\tau^2-\mpi^2}-\frac{3}{2}-\frac{\mpi^2 m_\tau^2\log\frac{\mpi^2}{m_\tau^2}}{(m_\tau^2-\mpi^2)^2}-8\pi^2 X_\ell(\mu_\text{UV})+\log\frac{\mu_\text{UV}^3}{\mpi^2 m_\tau}\notag\\
    &-2\bigg(1+\frac{m_\tau^2+\mpi^2}{m_\tau^2-\mpi^2}\log\frac{\mpi}{m_\tau}\bigg)\Big(\log\frac{\mu_\text{IR}^2}{\mpi m_\tau}-32\pi^2\Lambda_\text{IR}\Big)\Bigg\}\, ,
\label{Eq:matchingfs}
\end{align}
where $X_\ell(\mu)$ is defined as the combination
\begin{equation}
\label{RG_ChPT}
    X_\ell(\mu)=\frac{4}{3}X_1+X_6^{r}(\mu)-4K_{12}^r(\mu)\, ,\qquad 
 X_\ell(\mu)= X_\ell(M_\rho)+\frac{3}{16\pi^2}\log{\frac{\mu^2}{M_\rho^2}}\,.
\end{equation}
The long-distance part of this combination of LECs, $\bar X_\ell(M_\rho)$, was determined in Ref.~\cite{Ma:2021azh} by matching to lattice QCD~\cite{Feng:2020zdc}, yielding $\bar X_\ell(M_\rho)=14.0(6)\times 10^{-3}$, which differs quite significantly from the resonance-saturation estimate~\cite{Ananthanarayan:2004qk,Descotes-Genon:2005wrq,Bijnens:2014lea} 
\beq
\bar X_\ell(M_\rho)\simeq\frac{7}{32\pi^2}-\frac{3}{8\pi^2}\bigg(\frac{M_\rho}{4\pi F_\pi}\bigg)^2\simeq 5\times 10^{-3}\,.
\eeq
Moreover, so far we have glossed over the matching to SD contributions, which requires
\beq
\label{RG_LEFT}
\Delta X_\ell\big|_\text{SD}=-\frac{1}{4\pi^2}\log\frac{m_\tau^2}{M_\rho^2}
\eeq
to be added to $\bar X_\ell(\mu)$. This decomposition follows previous conventions for the SD contribution in the literature~\cite{Cirigliano:2001er,Cirigliano:2002pv,Davier:2002dy}. 
However, we emphasize that only the combination of $S_\text{EW}^{\pi\pi}$ and $G_\text{EM}(s)$ is scheme independent---at the precision at which the matching calculation is performed.\footnote{The scale $\mu_\text{UV}$ in the ChPT result cannot simply be identified with the scale of the low-energy effective (Fermi) theory after the $W$-boson is integrated out (LEFT), which is already reflected by the fact that the ChPT running~\eqref{RG_ChPT} and the LEFT running---Eq.~\eqref{RG_LEFT} arises from its naive application between $m_\tau$ and $M_\rho$---do not agree.
Instead, the matching between ChPT and LEFT is most conveniently formulated at the level of the LECs~\cite{Cirigliano:2023fnz}. For a consistent matching, the dependence on both the ChPT scale and the LEFT scale need to cancel in the decay rate at the considered order.} The resulting scheme ambiguity amounts to an $\Order(\alpha/\pi)$ uncertainty in $S_\text{EW}^{\pi\pi}$~\cite{Aliberti:2025beg}, which requires a dedicated matching calculation in analogy to Ref.~\cite{Cirigliano:2023fnz}, using input from lattice QCD for the required nonperturbative matrix elements~\cite{Feng:2020zdc,Yoo:2023gln}. Work along these lines is in progress~\cite{Cirigliano:2026ios}.

In this work, we take advantage of the fact that $\tilde f_+^{\text{ChPT}}(s,t)$, and consequently the final result $\tilde f_+^{\text{match}}(s,t)$, depend only linearly on $X_\ell(\mu)$, so that the dependence of $G_\text{EM}(s)$, given in Eq.~\eqref{eq:GEM} with Eqs.~\eqref{eq:Dst} and \eqref{eq:delta}, is also linear,
\begin{equation}
    \frac{\partial G_{\text{EM}}(s)}{\partial X_\ell(\mu)} = -e^2 \, ,
\end{equation}
and, at $\mathcal{O}(e^2)$, we can simply write 
\begin{equation}\label{eq:GemLECs}
    G_\text{EM}(s)\big|_{X_\ell(\mu)}
= G_\text{EM}(s) \big|_{X_\ell(\mu) = \bar{X}_\ell(\mu)} - e^2 \big[X_\ell(\mu) - \bar{X}_\ell(\mu)\big]\, .
\end{equation}
Our numerical results, presented for $\bar X_\ell(M_\rho)=14\times 10^{-3}$~\cite{Ma:2021azh}, are therefore trivial to adjust once an improved matching analysis becomes available. Our final result, $\tilde f_+^{\text{match}}(s,t)$ given in Eq.~\eqref{eq:f_match}, still involves the IR divergence, which only cancels once real-emission diagrams are included, to be addressed in the subsequent section.

\section{Real emission}
\label{sec:ReEm}

\begin{figure}[tb]
\centering
     \includegraphics[width=0.7\linewidth]{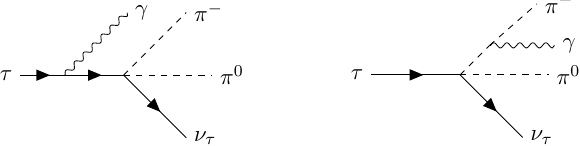}
     \caption{Initial- and final-state-radiations diagrams for $\tau^- \to \pi^- \pi^0 \nu_\tau$.}
    \label{Fig:ISR_FSR}
\end{figure}

The expression for the box diagram, Fig.~\ref{fig:loops}$(a)$, involves IR singularities that, together with the IR singularities in the remainder of the ChPT calculation, cancel once the initial- and final-state-radiation diagrams shown in Fig.~\ref{Fig:ISR_FSR} are included.  The matching procedure described in Sec.~\ref{sec:match} is therefore essential to ensure this cancellation also in the dispersive calculation, Fig.~\ref{fig:disp_tri}.

\begin{figure}[tb]
\centering
\includegraphics[width=0.9\linewidth]{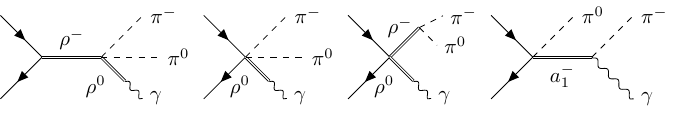}
\caption{$\rho$ and $a_1$ resonance-exchange contributions in resonance chiral theory~\cite{Ecker:1988te,Ecker:1989yg}.}
    \label{fig:chi_res}
\end{figure}

\begin{figure}[t]
    \centering
    \includegraphics[width=0.65\linewidth]{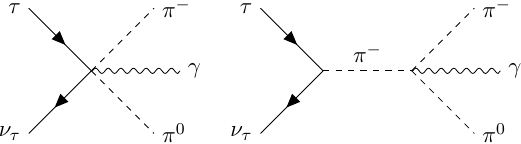}
    \caption{Chiral anomaly contribution via WZW action~\cite{Wess:1971yu,Witten:1983tw}.}
    \label{fig:WZW_res}
\end{figure}

Following Ref.~\cite{Cirigliano:2002pv}, the matrix element for the decay $\tau^-(l_1) \to \pi^-(q_1) \pi^0 (q_2) \nu_\tau (l_2)\gamma(k)$ has the general structure
\begin{align}\label{eq:amp_ReEm}
    T&=eG_FV_{ud}^*\epsilon^\mu(k)^*\big[F_\nu\bar{u}(l_2)\gamma^\nu(1-\gamma_5)(m_\tau+\slashed{l}_1-\slashed{k})\gamma_\mu u(l_1)\notag\\
    &\qquad +(V_{\mu\nu}-A_{\mu\nu})\bar{u}(l_2)\gamma^\nu(1-\gamma_5)u(l_1)\big]\, .
\end{align}
The first term in Eq.~\eqref{eq:amp_ReEm} describes bremsstrahlung off the initial $\tau$ lepton with
\begin{equation}
    F_\nu=\frac{(q_2-q_1)_\nu}{2l_1\cdot k}f_+(s)\, .
\end{equation}
The second part of the matrix element describes the vector and axial-vector components of the transition $W^-(l_1-l_2)\to \pi^-(q_1)\pi^0(q_2)\gamma(k)$. The hadronic tensor $V_{\mu\nu}$ contains bremsstrahlung off the $\pi^-$ in the final state and gauge invariance implies the Ward identities
\begin{equation}
    k^\mu V_{\mu\nu}=(q_1-q_2)_\nu f_+(s)\, ,\qquad
    k^\mu A_{\mu\nu}=0\, .
\end{equation}
For these resonance contributions, $V_{\mu\nu}$ and $A_{\mu\nu}$, we follow Ref.~\cite{Cirigliano:2002pv} and adopt
\begin{align}\label{eq:VmunuAmunu}
    V_{\mu\nu}&=f_+[(l_1-l_2)^2]\frac{q_{1\mu}}{q_1\cdot k}\left(q_1+k-q_2\right)_\nu-f_+[(l_1-l_2)^2]g_{\mu\nu}\notag\\
    &\qquad+\frac{f_+[(l_1-l_2)^2]-f_+(s)}{(q_1+q_2)\cdot k}\left(q_1+q_2\right)_\mu\left(q_2-q_1\right)_\nu\notag\\
    &\qquad +v_1\left(g_{\mu\nu}q_1\cdot k-q_{1\mu}k_\nu\right)+v_2\left(g_{\mu\nu}q_2\cdot k-q_{2\mu}k_\nu\right)\notag\\
    &\qquad +v_3\left(q_{1\mu}q_2\cdot k-q_{2\mu}q_1\cdot k\right)q_{1\nu}+v_4\left(q_{1\mu}q_2\cdot k-q_{2\mu}q_1\cdot k\right)\left(q_1+q_2+k\right)_\nu\, ,\notag\\
    A_{\mu\nu}&=ia_1\epsilon_{\mu\nu\rho\sigma}\left(q_2-q_1\right)^\rho k^\sigma+ia_2\left(l_1-l_2\right)_\nu\epsilon_{\mu\rho\sigma\tau}k^\rho q_1^\sigma q_2^\tau\, .
\end{align}
Taking into account $(l_1-l_2)^2=s+2(q_1+q_2)\cdot k$,  Low's theorem~\cite{Low:1958sn} is manifestly satisfied:
\begin{align}
    V_{\mu\nu}=&f_+(s)\frac{q_{1\mu}}{q_1\cdot k}\left(q_1-q_2\right)_\nu+f_+(s)\left(\frac{q_{1\mu}k_\nu}{q_1\cdot k}-g_{\mu\nu}\right)\notag\\
    &+2\frac{\diff f_+(s)}{\diff s}\left(\frac{q_{1\mu}q_2\cdot k}{q_1\cdot k}-q_{2\mu}\right)\left(q_1-q_2\right)_\nu+\mathcal{O}(k)\, . \label{Eq:Vmunu_expand}
\end{align}
Moreover, the vector components $v_{1\text{--}4}$ come from the resonance Lagrangian of Refs.~\cite{Ecker:1988te,Ecker:1989yg}, constructed in such a way that all resonance contributions that saturate the LECs $L_9$, $L_{10}$~\cite{Gasser:1983yg,Gasser:1984gg} are included, which leads to the diagrams listed in Fig.~\ref{fig:chi_res}. The axial-vector terms $a_{1,2}$ originate from the leading $\mathcal{O}(p^4)$ contribution of the Wess--Zumino--Witten (WZW) action~\cite{Wess:1971yu,Witten:1983tw} and the corresponding diagrams are the ones shown in Fig.~\ref{fig:WZW_res}. Since the strength of the WZW contribution is determined by the anomaly in terms of the pion decay constant, the free parameters in the resonance Lagrangian can be identified with two vector couplings $F_V$, $G_V$, one axial-vector coupling $F_A$, and the mass parameters $M_V$, $M_A$.

\begin{figure}
    \centering
		\includegraphics[width=0.45\linewidth]{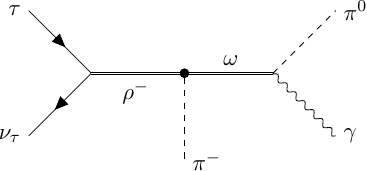}
    \caption{$\omega$ resonance contribution containing a radiative $\omega\pi\gamma$ coupling~\cite{Flores-Tlalpa:2005msx}.}
    \label{fig:omega_diagram}
\end{figure}

Additionally, we take into account the contribution from the resonance diagram containing a radiative $\omega\pi\gamma$ coupling, see Fig.~\ref{fig:omega_diagram}, since it has been noted before that it is sizable~\cite{Flores-Tlalpa:2005msx,Flores-Baez:2006yiq}, see Ref.~\cite{Flores-Tlalpa:2005msx} for explicit expressions. As in the analysis of $\tau^-\to\pi^-\pi^0\nu_\tau$ performed by the different experiments the decay $\omega\to\pi^0\gamma$ is considered a background contribution and excluded from the published data, we follow Refs.~\cite{Davier:2010fmf,Davier:2013sfa,Davier:2023fpl} in including only the interference contribution arising from the diagram with the radiative $\omega$ coupling.

To cast the real-emission contributions into the form anticipated in Eq.~\eqref{eq:delta} we now turn to the respective phase-space integrals. In general, we follow Refs.~\cite{Cirigliano:2001er,Cirigliano:2002pv} to calculate the fully inclusive decay rate, the exception concerning the IR singularities, which we treat in dimensional regularization instead of a photon-mass regulator.

\subsection[Determination of \texorpdfstring{$g_\text{Low}(s,t)$}{} and \texorpdfstring{$g_\text{rest}(s,t)$}{}]{Determination of \texorpdfstring{$\boldsymbol{g_\text{Low}(s,t)}$}{} and \texorpdfstring{$\boldsymbol{g_\text{rest}(s,t)}$}{}}

Once we consider only the leading Low approximation of $\mathcal{O}(k^{-2})$ to the differential decay rate, we get
\begin{equation}
    \text{d}\Gamma=\frac{\alpha G_F^2|V_{ud}|^2}{32\pi^7m_\tau}|f_+(s)|^2 D(s,t)\left[\frac{2l_1\cdot q_1}{(l_1\cdot k)(q_1\cdot k)}-\frac{m_\tau^2}{(l_1\cdot k)^2}-\frac{M_\pi^2}{(q_1\cdot k)^2}\right]\text{d}_\text{LIPS}\, ,
\end{equation}
with $D(s,t)$ from Eq.~\eqref{eq:Dst} in the isospin limit and where $d_\text{LIPS}$ denotes the four-body phase-space measure given by
\begin{equation}
    \text{d}_\text{LIPS}=\frac{\text{d}^3l_2}{2E_\nu}\frac{\text{d}^3q_1}{2E_{\pi^-}}\frac{\text{d}^3q_2}{2E_{\pi^0}}\frac{\text{d}^3k}{2k^0}\delta^{(4)}\big(l_1-l_2-q_1-q_2-k\big)\, ,
\end{equation}
and the kinematics of the radiative decay are discussed in more detail in App.~\ref{app:PhSpace}.
Integration over neutrino and photon momenta leads to the three-fold differential decay rate
\begin{align}
    \text{d}\Gamma&=\frac{\alpha G_F^2|V_{ud}|^2}{64\pi^4m_\tau^3}|f_+(s)|^2 D(s,t)\notag\\
    &\qquad\times\left[2l_1\cdot q_1I_{11}(s,t,x)-m_\tau^2I_{20}(s,t,x)-M_\pi^2I_{02}(s,t,x)\right]\text{d}s\,\text{d}t\,\text{d}x\, ,
\end{align}
with~\cite{Ginsberg:1967gvl}
\begin{equation}
    I_{mn}(s,t,x)=\frac{1}{2\pi}\int\frac{\text{d}^3l_2}{2E_\nu}\frac{\text{d}^3k}{2k^0}\,  \frac{\delta^{(4)}(l_1-l_2-q_1-q_2-k)}{(l_1\cdot k)^m(q_1\cdot k)^n}\, .
    \label{eq:Imn}
\end{equation}
The next step consists of performing the integration over $x$, the invariant mass squared of photon and neutrino. Here we must distinguish between two different regions in the $s$--$t$ plane. In the $(s,t)$-region of the Dalitz plot accessible to the nonradiative decay, the lower integration limit of Mandelstam $x$ is 0. For values of $(s,t)$ that cannot be accessed in the nonradiative decay the lower limit is given by $x_-(s,t)$ defined in App.~\ref{app:PhSpace}. The latter contribution is IR finite and occurs only for
\begin{equation}
    s\leq s_*\equiv\frac{m_\tau^2M_\pi}{m_\tau-M_\pi}\, .
\end{equation}
The corresponding contribution to $G_\text{EM}(s)$, booked in $g_\text{rest}(s,t)$, is enhanced in the threshold region and plays an important role in its contribution to $a_\mu$. The upper limit of the $x$ integration is always given by $x_+(s,t)$, as defined in App.~\ref{app:PhSpace}.

The double differential decay rate in the leading Low approximation, for $(s,t)$ in the nonradiative decay region integrated over the photon momentum, takes the form
\begin{equation}
    \text{d}\Gamma=\frac{\alpha G_F^2|V_{ud}|^2}{64\pi^4m_\tau^3}|f_+(s)|^2 D(s,t)\left[J_{11}(s,t)+J_{20}(s,t)+J_{02}(s,t)\right]\text{d}s\,\text{d}t\, ,
\end{equation}
where
\begin{align}
    J_{11}(s,t)&=\bigg(32\pi^2\Lambda_\text{IR}+1-\log\frac{\mu_\text{IR}^2}{[2x_+(s,t)\bar{\gamma}]^2}\bigg)\frac{t(\mpi^2+m_\tau^2-t) B_0^{\pi\tau}(t)}{\lambda_{\pi\tau}(t)}\notag\\
    &+\frac{1}{\bar{\beta}}\left\{\text{Li}_2\left(\frac{1}{Y_2}\right)-\text{Li}_2\left(Y_1\right)+\frac{1}{4}\left[\log^2\left(-\frac{1}{Y_2}\right)-\log^2\left(-\frac{1}{Y_1}\right)\right]\right\}\,,\notag\\
    J_{20}(s,t)&=-\frac{1}{2}\bigg(32\pi^2\Lambda_\text{IR}+1-\log\frac{\mu_\text{IR}^2}{m_\tau^2}\bigg)+\log\frac{m_\tau^2-s}{x_+(s,t)}\,,\notag\\
    J_{02}(s,t)&=-\frac{1}{2}\bigg(32\pi^2\Lambda_\text{IR}+1-\log\frac{\mu_\text{IR}^2}{\mpi^2}\bigg)+\log\frac{m_\tau^2+M_\pi^2-s-t}{x_+(s,t)}\, ,
    \label{eq:Js}
\end{align}
are the functions worked out in Ref.~\cite{Cirigliano:2002pv} and defined in terms of the quantities reported in App.~\ref{app:PhSpace}, but with the IR-divergent parts translated to dimensional regularization. Thus, $g_\text{Low}(s,t)$ is given by
\begin{equation}\label{eq:gbrem}
    g_\text{Low}(s,t)=\frac{\alpha}{\pi}\left[J_{11}(s,t)+J_{20}(s,t)+J_{02}(s,t)\right]\, .
\end{equation}
Added to the virtual contributions, see Eq.~\eqref{eq:delta}, 
the IR divergent piece in $J_{11}(s,t)$ cancels exactly with the IR divergence in $\tilde f_+^{(a)}(s,t)$, and likewise for $J_{20}(s,t)+J_{02}(s,t)$ and $\tilde f_+^{\text{CT}+\text{SE}}(s,t)$. By virtue of the matching in Sec.~\ref{sec:match}, the same cancellation applies to our final $\tilde f_+^\text{match}(s,t)$. 

The remaining part of the radiative decay rate, containing all terms except for the leading Low ones, is calculated numerically and is collected in $g_\text{rest}(s,t)$, which accounts for the IR-finite remainder of the rate, i.e., all terms except for the ones contained in $g_\text{Low}(s,t)$.

\subsection[Soft-photon limit in \texorpdfstring{$g_\text{rest}(s,t)$}{}]{Soft-photon limit in \texorpdfstring{$\boldsymbol{g_\text{rest}(s,t)}$}{}}

As remarked earlier, we compute the contribution of $g_\text{rest}(s,t)$ purely numerically. This is done by integrating the square of the amplitude $T$ given in Eq.~\eqref{eq:amp_ReEm}. However, the piece that gives rise to the IR-divergent $g_\text{Low}(s,t)$ needs to be subtracted. Hence, we write $g_\text{rest}(s,t)$ as
\begin{align}
\label{Eq:g_rest}
    g_\text{rest}(s,t) &= \frac{1}{128 \pi^3 G_F^2 |V_{ud}|^2 D(s,t) |f_+(s)|^2}\\
    &\times \int \diff x\, \diff \text{cos}\, \theta_k\, \diff \phi_k\, \frac{m_\tau^2 x}{\left(m_\tau^2-s+x+\cos\theta_k\sqrt{\lambda\left(s,x,m_\tau^2\right)}\right)^2} \sum\limits_{\text{spins}} \Big\{ |T|^2 - |T_\text{Low}|^2\Big\} \, , \notag
\end{align}
with the measure function of the four-body phase-space integral spelled out explicitly, see Eq.~\eqref{Eq:meas} in App.~\ref{app:PhSpace}, and the amplitude square $|T_\text{Low}|^2$. The latter gives rise to Eq.~\eqref{eq:Js}, once expressed as
\begin{equation}
    \sum\limits_{\text{spins}}|T_\text{Low}|^2 = 8 e^2 G_F^2 |V_{ud}|^2 |f_+(s)|^2 D(s,t) \left[\frac{2l_1\cdot q_1}{(l_1\cdot k)(q_1\cdot k)}-\frac{m_\tau^2}{(l_1\cdot k)^2}-\frac{M_\pi^2}{(q_1\cdot k)^2}\right]\, ,
\end{equation}
and the subtraction in Eq.~\eqref{Eq:g_rest} yields an expression for $g_\text{rest}(s,t)$ free of IR singularities.

\subsection{Integration at threshold}

In the numerical integration of $g_\text{Low}(s,t)$ in $t$ and $g_\text{rest}(s,t)$ in $\phi_k$, $\cos\theta_k$, $x$, and $t$ (for details about the phase-space integration see App.~\ref{app:PhSpace}) a divergence of the integrand at the two-pion threshold $s=4M_\pi^2$ appears. In order to bypass this endpoint singularity we apply a change of variables where we parameterize $t$ and $x$ in terms of suitably chosen angles:
\begin{align}
    x&\rightarrow\begin{cases}
        \frac{1}{2}x_+(s,t)\left(z_x-1\right) &\text{in}\quad R^\text{III}\, ,\\
         \frac{1}{2}\left[x_+(s,t)-x_-(s,t)\right]z_x+\frac{1}{2}\left[x_+(s,t)+x_-(s,t)\right] &\text{in}\quad R^\text{IV}\backslash R^\text{III}\, ,
    \end{cases}\notag\\
    t&\rightarrow\begin{cases}
        \frac{1}{2}\left[\bar{t}_\text{max}(s)-\bar{t}_\text{min}(s)\right]z_t+\frac{1}{2}\left[\bar{t}_\text{max}(s)+\bar{t}_\text{min}(s)\right] &\text{in}\quad R^\text{III}\,,\\
         \frac{1}{2}\left[(m_\tau-M_\pi)^2-\bar{t}_\text{max}(s)\right]z_t+\frac{1}{2}\left[(m_\tau-M_\pi)^2+\bar{t}_\text{max}(s)\right]&\text{in}\quad R^\text{IV}\backslash R^\text{III}\,,
    \end{cases}
    \label{eq:t_ang}
\end{align}
where $-1\leq z_x\leq1$ and $-1\leq z_t\leq1$ and the expressions for $x_\pm(s,t)$ and $\bar{t}_{\text{min,max}}(s)$ are given in App.~\ref{app:PhSpace}. Here, $R^\text{IV}$ refers to the $(s,t)$-region in the Dalitz plot accessible to the radiative decay, while $R^\text{III}$ refers to the region accessible to the nonradiative decay. With this transformation we can isolate the threshold divergence and numerically integrate the amplitude over the full $s$ range, $s\in[4M_\pi^2,m_\tau^2]$. In particular, it is the product of the radiative amplitude (after the change of variables) times the Jacobian arising from rewriting $t$ in terms of $z_t$ that becomes integrable at threshold. In fact, it is easy to see from Eq.~\eqref{eq:t_ang} that the Jacobian in $R^\text{III}$, where the singular behavior arises, reads
\begin{equation}
    J_t(s)=\frac{1}{2}\left[\bar{t}_\text{max}(s)-\bar{t}_\text{min}(s)\right]\,,
\end{equation}
and it vanishes at $s=4M_\pi^2$ since $\bar{t}_\text{max}(s)=\bar{t}_\text{min}(s)$, curing the divergence. The Jacobian in $R^\text{IV}\backslash R^\text{III}$, instead, does not vanish at threshold, and this results in a regular but sizable contribution of the real-emission diagrams in this region.
After testing this approach with the real-emission contributions, we applied it to the determination of $\tilde f_+^\text{disp}(s,t)$ as well,  so that also in this case the integration remains under control down to the two-pion threshold.

\subsection[Threshold singularity of \texorpdfstring{$g_\text{Low}(s,t)$}{}]{Threshold singularity of \texorpdfstring{$\boldsymbol{g_\text{Low}(s,t)}$}{}}
\label{sec:threshold_gLow}

The variable substitution described in the previous section proves valuable for the integration of $\tilde f_+^\text{match}(s,t)$ and $g_\text{rest}(s,t)$, but it does not work for $g_\text{Low}(s,t)$ defined in Eq.~\eqref{eq:gbrem}. In fact, this result shows a physical divergence at threshold, $s=4M_\pi^2$, which is not solved by rewriting the variable $t$ in terms of the angle $z_t$ and then multiplying with the Jacobian. 

In order to solve this problem, we performed the variable substitution also in this case, but then expanded the new expression around $s=4M_\pi^2$. The contribution to $G_\text{EM}(s)$ from $g_\text{Low}(s,z_t)$, after the angular integration, i.e.,
\begin{align}
    G_\text{EM}^J(s)&=\frac{\int_{-1}^1\text{d}z_t\,J_t(s)D(s,z_t)g_\text{Low}(s,z_t)}{\int_{-1}^1\text{d}z_t\, J_t(s) D(s,z_t)}\, ,\label{eq:Gembrem}\\ 
 \int_{-1}^1\text{d}z_t\, J_t(s) D(s,z_t)&=\frac{m^2}{6s}(s-m_\tau^2)^2(2s+m_\tau^2)(s-4M_\pi^2)^{3/2}\,,\notag
\end{align}
has the following analytical form
\begin{align}
    G_\text{EM}^J(s)&= \left[c^\text{log}_{3/2}\log\left(\frac{s-4M_\pi^2}{4M_\pi^2}\right)+c^\text{rest}_{3/2}\right]+c^\text{rest}_{2}\sqrt{s-4M_\pi^2}\notag\\
    &+(s-4M_\pi^2)\left[c^\text{log}_{5/2}\log\left(\frac{s-4M_\pi^2}{4M_\pi^2}\right)+c^\text{rest}_{5/2}\right]+\mathcal{O}\big[(s-4M_\pi^2)^{3/2}\big]\, ,
\end{align}
where the $c_i^\text{rest}$ include all the terms that are not divergent for $s=4M_\pi^2$, with $i=3/2,2,5/2,\ldots$ indicating the degree of the expansion of the numerator in Eq.~\eqref{eq:Gembrem}. It is then clear that, after the integration in $z_t$, the contribution of $g_\text{Low}(s,z_t)$ to $G_\text{EM}(s)$ is still logarithmically divergent at threshold.

\subsection{Determination of resonance couplings}
\label{sec:newFA}

For the numerical evaluation of the resonance contribution to real emission we need to determine the couplings $F_V$, $G_V$, and $F_A$. A frequent choice relies on SD constraints~\cite{Ecker:1989yg}
    \begin{equation}\label{eq:par_SD}
        F_V=\sqrt{2}F_\pi\simeq 0.13\GeV\, ,\qquad G_V=\frac{F_\pi}{\sqrt{2}}\simeq 0.065\GeV\, ,\qquad F_A=F_\pi\simeq 0.092\GeV\, ,
    \end{equation}
which we will consider as one parameter set. Alternatively, phenomenological analyses suggest
    \begin{equation}\label{eq:par_ph}
        F_V=0.16\GeV\, ,\qquad G_V=0.065\GeV\, ,\qquad F_A=0.12\GeV\, ,
    \end{equation}
    where $F_V$ and $G_V$ are extracted from the decay widths of $\rho\to e^+e^-$, $\rho\to\pi\pi$, and $K^*\to K\pi$~\cite{Colangelo:2021moe}, while $F_A$ originates from a measurement  of the $a_1\to\pi\gamma$ partial width~\cite{Zielinski:1984au}. The latter extraction of $F_A$ from $a_1\to\pi\gamma$ has drawn criticism in the literature, with alternative strategies mostly suggesting smaller values of $F_A$ \cite{Garcia-Martin:2010kyn,Knecht:2001xc,Moussallam:1997xx,Cirigliano:2004ue,Condo:1993xa,CLAS:2008zko,Lueghausen:2019}. Given that the situation is far from conclusive, we attempt another indirect determination here, by assuming that the $a_1\to\pi\gamma$ decay proceeds via $a_1\to\rho\pi\to\pi\gamma$~\cite{Hoferichter:2017ftn,Zanke:2021wiq}.

For the first step of the decay chain $a_1 \to \pi \rho \to \pi \gamma$, one needs the coupling for $a_1\to \rho\pi$, which could be defined in a  hidden-local-symmetry model~\cite{Bando:1987br} according to
	\begin{equation}
		\mathcal{L}_{a_1 \pi \rho} = - \frac{i}{\sqrt{2}} g_{a_1 \rho \pi} \operatorname{Tr} \mathcal{A}_{\mu} [\Phi,\, V^{\mu}]\,,
	\end{equation}
	where the relevant fields are collected in
	\begin{equation}
		\Phi =
		\begin{pmatrix}
			\pi^0/\sqrt{2} & \pi^+\\
			\pi^- &-\pi^0/\sqrt{2}
		\end{pmatrix}\,,\quad
		V_{\mu} =
		\begin{pmatrix}
			\rho^0/\sqrt{2} & \rho^+\\
			\rho^- &-\rho^0/\sqrt{2}
		\end{pmatrix}_{\mu}\,,\quad
		\mathcal{A}_{\mu}=
		\begin{pmatrix}
			a_1^0/\sqrt{2} & a_1^+\\
			a_1^- &-a_1^0/\sqrt{2}
		\end{pmatrix}_{\mu}\,.
	\end{equation}
	However, in order to ensure gauge invariance, we promote it to
	\begin{equation}
		\tilde{\mathcal{L}}_{a_1 \pi \rho} = - \frac{i}{2\sqrt{2}} \tilde{g}_{a_1 \rho \pi} \operatorname{Tr} \mathcal{A}_{\mu \nu} [\Phi,\, V^{\mu \nu}]\,,
	\end{equation}
	formulated in terms of the field-strength tensors $V_{\mu \nu} = \partial_\mu V_\nu - \partial_\nu V_\mu$ and $\mathcal{A}_{\mu \nu} = \partial_\mu \mathcal{A}_\nu - \partial_\nu \mathcal{A}_\mu$. Momentum-space Feynman rules extracted from this interaction term appear as, e.g.,
	\begin{equation}
	\vcenter{\hbox{\includegraphics[width=0.4\linewidth]{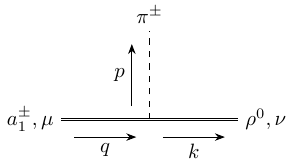}}}
		= \pm \tilde{g}_{a_1 \rho \pi} (g_{\mu \nu}\,k\cdot q - k_\mu q_\nu)\,,
	\end{equation}
	similarly to the $a_1 \rho \pi$ interaction term utilized in Refs.~\cite{LopezCastro:1999xg,Flores-Tlalpa:2005msx,Flores-Baez:2006yiq}.
	Since the interactions for the different channels mediated by the above interaction term differ only by a sign, the partial widths appear as
	\begin{equation}
		\Gamma(a_2^\pm \to \pi^\pm \rho^0) = \Gamma(a_2^\pm \to \pi^0 \rho^\pm) = \Gamma(a_1^0 \to \pi^\pm \rho^\mp)\,.
	\end{equation}
	We can express
	\begin{equation}
		\Gamma(a_1 \to \pi \rho) = 2\Gamma(a_2^\pm \to \pi^\pm \rho^0)=\frac{|\tilde{g}_{a_1 \rho \pi}|^2}{4 \pi}  M_\rho^2 |\mathbf{p}_\rho| \bigg(1 + \frac{2 |\mathbf{p}_\rho|^2}{3M_\rho^2}\bigg)\,,
	\end{equation}
	with the spatial $\rho$ momentum $|\mathbf{p}_\rho| = \sqrt{\lambda(M_{a_1}^2,M_\rho^2,M_\pi^2)}/(2M_{a_1})$ in the center-of-mass frame.
	Further assuming the total experimental width, $\Gamma_{a_1} = (0.25,\ldots,0.6)\GeV$~\cite{ParticleDataGroup:2024cfk}, to be dominated by the $\pi \rho$ decay channel, $\Gamma_{a_1} = \Gamma(a_1 \to \pi \rho)$, yields
	\begin{equation}
    \label{ga1rhopi}
		|\tilde{g}_{a_1 \rho \pi}| = (3.6,\ldots,5.6) \GeV^{-1}\,,
	\end{equation}
	or $|g_{a_1 \rho \pi}| = 5.1\GeV^{-1}$ corresponding to $\Gamma_{a_1} = 0.5\GeV$ used in the evaluation of the resonance terms of Eq.~\eqref{eq:VmunuAmunu}. The total width as measured by the COMPASS collaboration $\Gamma_{a_1} = 0.38(8)\GeV$~\cite{COMPASS:2018uzl} leads to a coupling in the middle of this range, $|g_{a_1 \rho \pi}| = 4.5(5)\GeV^{-1}$, and also the coupling $g_{\rho a_1 \pi}=4.843 \GeV^{-1}$ used in Ref.~\cite{Flores-Tlalpa:2005msx} falls within the range given in Eq.~\eqref{ga1rhopi}.
    
	In order to describe the full decay chain, we make further use of the gauge-invariant interaction term of Ref.~\cite{Klingl:1996by},
	\begin{equation}
		\mathcal{L}_{\gamma V} = \frac{\sqrt{2} e}{g_{\rho \gamma}} F^{\mu \nu} \operatorname{Tr} Q V_{\mu \nu}\,,
	\end{equation}
	with the charge matrix $Q = \operatorname{diag}(2,-1)/3$ and the electromagnetic field-strength tensors $F_{\mu \nu} = \partial_\mu A_\nu - \partial_\nu A_\nu$. Coupling the photon to a neutral $\rho$ with momentum $k$, the partial decay width can be written as
	\begin{equation}
		\Gamma(a_1^\pm \to \pi^\pm \gamma) = \frac{e^2 |\tilde{g}_{a_1 \rho \pi}|^2}{24 \pi g_{\rho \gamma}^2}  \bigg(\frac{k^2}{M_{\rho}^2 - k^2}\bigg)^2 |\mathbf{k}| \bigg(3 k^2 + 2|\mathbf{k}|^2 \bigg)\,,
	\end{equation}
	where the spatial photon momentum is given by $|\mathbf{k}|=\sqrt{\lambda(M_{a_1}^2,M_\pi^2,k^2)}/(2 M_{a_1})$. Taking the limits $M_\rho^2\to 0$, $k^2 \to 0$, the above expression yields
	\begin{equation}
		\Gamma(a_1^\pm \to \pi^\pm \gamma) = \frac{e^2 |\tilde{g}_{a_1 \rho \pi}|^2 M_{a_1}^3}{96 \pi g_{\rho\gamma}^2} \bigg(1 - \frac{M_\pi^2}{M_{a_1}^2} \bigg)^3\,.
	\end{equation}
    Equating this result with~\cite{Ecker:1988te}
	\begin{equation}
		\label{Eq:Gam_FA}
		\Gamma(a_1^\pm \to \pi^\pm \gamma) = \frac{e^2 F_A^2 M_{a_1}}{96 \pi F_\pi^2} \bigg(1 - \frac{M_\pi^2}{M_{a_1}^2} \bigg)^3\,,
	\end{equation}
	we obtain
	\begin{equation}
		F_A = \{0.069,\, 0.097,\, 0.11\} \GeV\,,
	\end{equation}
	for $\Gamma_{a_1} = \{0.25,\, 0.5,\, 0.6\} \GeV$, where $g_{\rho \gamma} = 5.98$~\cite{Holz:2024diw} was employed. With the total width as measured by COMPASS~\cite{COMPASS:2018uzl} one would obtain $F_A = 0.085(9) \GeV$. Further ambiguity in the numerical evaluation arises when using $g_{\rho \gamma} = 4.96$ as extracted from the analytic continuation of the pion VFF~\cite{Zanke:2021wiq,Hoferichter:2023mgy},
	\begin{equation}
		F_A = \{0.083,\, 0.12,\, 0.13\} \GeV\,,
	\end{equation}
	or $F_A = 0.10(1) \GeV$ using the COMPASS total $a_1$ width in this case. Altogether, depending on the assumptions for the $a_1$ parameters, we end up with a range 
    \beq
    \label{FA_our}
    F_A=(0.07,\ldots,0.13)\GeV\,.
    \eeq
    For the numerical analysis, we will use the average of results obtained from the parameter sets~\eqref{eq:par_SD} and~\eqref{eq:par_ph}, with their difference interpreted as a $1\sigma$ error. For $F_A$, this amounts to $F_A=0.106(28)\GeV$, in reasonable agreement with the range~\eqref{FA_our} indirectly obtained from $a_1 \to \pi \rho \to \pi \gamma$.

\section{Fits to the \texorpdfstring{$\boldsymbol{\tau}$}{} spectral function}
\label{sec:fits}

In principle, we now have all the ingredients to compute $G_\text{EM}(s)$~\eqref{eq:GEM}, replacing the ChPT definition~\eqref{eq:delta} by our full result
\begin{equation}
    \Delta(s,t)=1+2\,\Re\tilde f_+^{\text{match}}(s,t)+g_{\text{Low}}(s,t)+g_{\text{rest}}(s,t)\,.
\label{eq:delta_match}
\end{equation}
However, to determine the input for $\Im f_+(s)$ in the evaluation of the dispersive integrals in a self-consistent way, we first need to consider fits to the $\tau$ spectral function~\eqref{decay_rate}.

\subsection{Fitting procedure}

The pion VFF $f_+(s)$ is determined from the spectra measured by Belle~\cite{Belle:2008xpe}, ALEPH~\cite{ALEPH:2005qgp,Davier:2013sfa}, CLEO~\cite{CLEO:1999dln}, and OPAL~\cite{OPAL:1998rrm}, providing data for 
\begin{equation}\label{eq:Belle_spectrum}
    \frac{1}{N}\frac{\text{d}N}{\text{d}s}=\frac{K_\Gamma(s)}{\Gamma_\pi}\big[\beta_{\pi\pi^0}(s)\big]^3|f_+(s)|^2G_\text{EM}(s)\, ,
\end{equation}
with partial widths $\Gamma_e\equiv\Gamma(\tau^-\to e^-\bar{\nu}_e\nu_\tau)$, contained in $K_\Gamma(s)$~\eqref{decay_rate}, and $\Gamma_\pi\equiv\Gamma(\tau^-\to\pi^-\pi^0\nu_\tau)$. 
For the external inputs we use $S_\text{EW}^{\pi\pi}=1.0233(24)$~\cite{Aliberti:2025beg},  $\Br[\tau\to e\nu_\tau\bar{\nu}_e]=17.82(4)\%$ and $\Br[\tau\to \pi\pi\nu_\tau]=25.49(9)\%$ from the global fit of Refs.~\cite{ParticleDataGroup:2024cfk,HeavyFlavorAveragingGroupHFLAV:2024ctg} (whose errors are correlated with coefficient $-0.19$), and $V_{ud}=0.97367(32)$~\cite{ParticleDataGroup:2024cfk}. 

Since $G_\text{EM}(s)$ itself depends on the input of $f_+(s)$, but is rather expensive to  calculate numerically, we apply an iterative procedure:
\begin{enumerate}
    \item First, we determine $G_\text{EM}(s)$ from $f_+(s)=\Omega^1_1(s)$, with $\delta^1_1(s_0)=110.4^\circ$ and $\delta^1_1(s_1)=165.7^\circ$~\cite{Colangelo:2018mtw}.
    \item Having calculated this first approximation for $G_\text{EM}(s)$, we fit the free parameters in $f_+(s)$ of Eq.~\eqref{eq:piFF} by means of Eq.~\eqref{eq:Belle_spectrum} to the experimental data, using the mapping~\eqref{eq:smap} to ensure that the threshold lies at its physical value.
    \item The resulting representation for $f_+(s)$ is then used to calculate $G_\text{EM}(s)$ in the next step and the fit is repeated with this new input.
\end{enumerate}
The procedure converges after a few iterations. In practice, the differences to the starting point $f_+(s)=\Omega^1_1(s)$ are small but certainly not negligible, reflecting the important role of inelastic effects for a precision analysis of $a_\mu^\text{HVP, LO}[\pi\pi]$~\cite{Chanturia:2022rcz,Heuser:2024biq}.

For the practical implementation, we need to account for the finite bin size $s_b$, which especially in the vicinity of resonances can have a major impact, e.g., for the Belle data~\cite{Belle:2008xpe} almost all of the bins are of width $s_i^b=0.05\GeV^2$. Within the bin, events will not be distributed equally, but weighted by the distribution, so that the actual observable becomes 
\begin{equation}
    \left(\frac{1}{N}\frac{\text{d}N}{\text{d}s}(s_i)\right)^{\text{bin}}=\frac{1}{s_i^b}\int^{s_i+s_i^b/2}_{s_i-s_i^b/2}\text{d}s\, \frac{1}{N}\frac{\text{d}N}{\text{d}s}(s)\, ,
\end{equation}
for the bins with center $s_i$. Alternatively, one can calculate corrected bin centers $s_i^\text{corr}$ for the pion VFF fit by a null search, and we checked that both approaches lead to identical results. This effect is analogous to the required bin average for hadronic cross-section measurements using initial-state radiation~\cite{Colangelo:2018mtw,Stamen:2022uqh,Hoferichter:2023bjm}.  

Furthermore, we make use of the statistical covariance matrix $\text{Cov}^\text{stat}(i,j)$ as well as the full systematic covariance matrix $\text{Cov}^\text{syst}(i,j)$ whenever available, which is the case for the Belle and ALEPH experiments. While for the CLEO experiment only the systematic covariance matrix is available, the OPAL experiment provides a combined statistical and systematic covariance matrix, without further information to disentangle the two error components. In these two cases we treated the single covariance matrix as purely statistical. In order to avoid the D'Agostini bias~\cite{DAgostini:1993arp}, the total covariance matrix is constructed via~\cite{Ball:2009qv}
\begin{equation}
    \text{Cov}(i,j)=\text{Cov}^\text{stat}(i,j)+\left(\frac{1}{N}\frac{\text{d}N}{\text{d}s}(s_i)\right)^\text{bin}\left(\frac{1}{N}\frac{\text{d}N}{\text{d}s}(s_j)\right)^\text{bin}\frac{\text{Cov}^\text{syst}(i,j)}{y_iy_j}\, ,
\end{equation}
where $y_i$ is the central value of the experimental observable in bin $i$, which is then used in yet another iterative procedure for each fit of the theoretical distribution to the data.

\subsection{Sources of uncertainties}
\label{sec:err}

The uncertainties we considered in our analysis and that we propagated to $G_\text{EM}(s)$ can be divided into the ones coming from the input data and the theoretical ones. For the first group, the uncertainties are given by
\begin{enumerate}
    \item statistical and systematic covariance matrices of the $\tau^-\to\pi^-\pi^0\nu_\tau$ data,
    \item uncertainties on the spectral function due to the branching ratios $\text{Br}(\tau^-\to e^-\bar{\nu}_e\nu_\tau)$ and $\text{Br}(\tau^-\to \pi^-\pi^0\nu_\tau)$.
\end{enumerate}
The errors derived from the covariance matrices
give rise to a fit uncertainty on $f_+(s)$, which is then propagated to $G_\text{EM}(s)$, including a scale factor if $\chi^2/\text{dof}>1$. Combined with the uncertainty derived from the branching ratios, this defines the experimental error. 
On the theory side, the sources of uncertainties entering in the pion VFF $f_+(s)$ are
\begin{enumerate}
    \item the variation of the conformal polynomial degree $N$,
    \item the variation of the $s_c$ parameter in the conformal polynomial,
\end{enumerate}
while for uncertainties entering $G_\text{EM}(s)$ directly we have
\begin{enumerate}
\setcounter{enumi}{2} 
    \item the variation of the cutoff of the dispersive integral (from 20 $\text{GeV}^2$ to 9 $\text{GeV}^2$),
    \item the use of different estimates for the couplings $F_V$, $G_V$, and $F_A$, which enter in the resonance contribution to real emission, see Sec.~\ref{sec:newFA}.   
\end{enumerate}
In the latter case, we define the difference between results obtained with the SD couplings~\eqref{eq:par_SD} and the phenomenologically estimated ones~\eqref{eq:par_ph} as $1\sigma$ uncertainties, while for the other uncertainties the errors are quoted as the maximal deviation from our central solution. 

\begin{table}[t]
\renewcommand{\arraystretch}{1.3}
    \centering
    \adjustbox{width = \textwidth}{\begin{tabular}{l r r r r c c r r r r r r}
				\toprule
				 $\chi^2/\text{dof}$ & $p$-value & $\delta(s_0)$ $[^\circ]$ & $\delta(s_1)$ $[^\circ]$ & $p_3\times10^2$ & $p_4\times10^2$ & $p_5\times10^2$ &  $c_{\rho'}\ [\text{GeV}^2]$ & $M_{\rho'}\ [\text{GeV}]$& $\Gamma_{\rho'}\ [\text{MeV}]$ & $c_{\rho''}\ [\text{GeV}^2]$ & $M_{\rho''}\ [\text{GeV}]$ & $\Gamma_{\rho''}\ [\text{MeV}]$ \\ \midrule
				 \multicolumn{13}{c}{Belle ($N_\text{data}=62$)}\\\midrule
				 $1.36$ &$0.041$ & $110.05(13)$ & $167.00(10)$ & $3.11(65)$ & & & $0.47(9)$ & $1.46(2)$ & $443(47)$ & $-0.24(8)$ & $1.67(2)$ & $257(83)$\\
				 $1.30$ &$0.070$ & $109.95(15)$ & $166.93(11)$ & $-3.7(4.4)$ & $3.4(2.2)$ & & $0.33(10)$ & $1.44(2)$ & $375(58)$ & $-0.40(14)$ & $1.70(2)$ & $289(76)$\\
				 $0.93$ &$0.62$ & $109.60(16)$ & $166.57(13)$ & $-4.3(4.7)$ & $17.5(3.7)$ & $-6.4(1.4)$ & $0.66(12)$ & $1.49(1)$ & $566(53) $& $-0.94(20)$ & $1.75(1)$ & $398(58)$\\ \midrule
				 $1.36$ &$0.043$ &$ 110.04(14)$ & $167.00(10)$ & $3.13(66)$ & & & $0.47(10)$ & $1.46(2)$ & $442(51)$ & $-0.24(8)$ & $1.67(2)$ & $257(87)$\\
				 $1.30$ &$0.072$ & $109.94(15)$ & $166.92(11)$ & $-3.6(4.4)$ & $3.3(2.2)$ & & $0.33(10)$ & $1.44(2)$ & $376(58)$ & $-0.39(14)$ &$ 1.70(2) $& $289(77)$\\
				 $0.93$ &$0.62$ & $109.59(26)$ & $166.57(26)$ & $-4.2(4.9)$ & $17.4(5.5)$ & $-6.4(2.1)$ & $0.66(12)$ & $1.49(1)$ & $565(57)$ & $-0.94(21) $& $1.75(1)$ & $399(63)$ \\ \midrule
                 \multicolumn{13}{c}{Belle+ALEPH ($N_\text{data}=62+78=140$)}\\\midrule
				 $1.37$ &$0.0029$ & $109.89(13)$ & $166.77(8)$ & $3.11(64)$ & & & $0.46(8)$ & $1.46(2)$ & $450(47)$ & $-0.21(7)$ & $1.67(2)$ & $262(98)$\\
				 $1.38$ &$0.0028$ & $109.81(13)$ & $166.71(10)$ & $-1.8(3.1)$ & $2.4(1.6)$ & & $0.36(6)$ &$ 1.45(1)$ & $404(35)$ & $-0.32(10)$ & $1.70(2)$ & $297(71)$\\
				 $1.20$ &$0.062$ & $109.41(14)$ & $166.33(10)$ & $-0.77(4.80)$ & $17.7(3.4)$ & $-7.3(1.4)$ & $0.75(13)$ & $1.50(1)$ & $611(52)$ & $-0.82(19)$ & $1.75(1) $& $400(59)$\\ \midrule
				 $1.37$ &$0.0030 $& $109.88(12)$ & $166.77(9)$ & $3.13(65)$ & & &$ 0.46(9)$ & $1.46(2)$ & $450(47)$ & $-0.21(8)$ & $1.67(2)$ & $260(120)$\\
				 $1.38$ &$0.0030$ & $109.79(14)$ & $166.71(10)$ & $-1.7(4.4)$ & $2.4(2.2)$ & & $0.37(10)$ & $1.45(2)$ & $405(57) $& $-0.32(14)$ & $1.70(3)$ & $297(92)$\\
				 $1.20 $&$0.060 $& $109.41(13)$ & $166.33(9)$ & $-0.7(4.8)$ & $17.5(3.2)$ & $-7.2(1.3)$ &$ 0.75(13)$ & $1.50(1)$ & $610(51)$ & $-0.81(19)$ & $1.75(1)$ & $401(59)$\\ \midrule
				 \multicolumn{13}{c}{Belle+ALEPH+CLEO+OPAL ($N_\text{data}=62+78+43+72=255$)} \\ \midrule
				 $1.32$ &$0.0006$ & $109.75(9)$ & $166.59(7)$ & $3.26(69)$ & & & $0.53(10)$ & $1.47(2)$ & $482(49) $& $-0.25(8)$ & $1.66(2)$ & $295(69)$\\
				 $1.32 $&$0.0006$ & $109.67(12)$ & $166.54(9)$ & $-1.4(4.2)$ & $2.2(2.1)$ & & $0.43(12)$ & $1.46(2) $& $439(57)$ & $-0.36(14)$ & $1.69(3)$ & $327(90)$\\
				 $1.19$ &$0.021 $& $109.31(12)$ & $166.20(9)$ & $-0.8(5.9)$ & $18.6(3.2)$ & $-7.7(1.4)$ & $0.81(21)$ & $1.50(2)$ &$ 631(75) $& $-0.88(19)$ & $1.75(1)$ & $416(57)$\\ \midrule
				 $1.32$ &$0.0006$ & $109.74(11)$ & $166.59(8)$ & $3.28(67)$ & & & $0.53(8)$ & $1.47(2)$ & $481(44)$ & $-0.25(5)$ & $1.66(2)$ & $295(52)$\\
				 $1.32$ &$0.0006$& $109.66(12)$ & $166.54(9)$ & $-1.2(3.8)$ & $2.2(1.9)$ & & $0.43(10)$ & $1.46(2)$ & $440(50)$ & $-0.36(13)$ & $1.69(3)$ & $326(85)$\\
				 $1.19 $&$0.020 $& $109.31(15) $& $166.18(14)$ & $-0.7(5.0)$ &$ 18.4(4.0)$ & $-7.6(1.7)$ & $0.81(14) $& $1.50(1) $& $630(59) $& $-0.88(20) $& $1.75(1)$ & $417(68)$\\ \bottomrule
                 \renewcommand{\arraystretch}{1.0}
			\end{tabular}
            }

    \caption{Results of our fits to the (combined) data sets Belle, Belle+ALEPH, and Belle+ALEPH+CLEO+OPAL (in brackets the number of data points). For each (combination of) experimental data set(s), we show the result for $N=3$ (first line of each set of fits), $N=4$ (second line of each set of fits), and $N=5$ (third line of each set of fits). Moreover, the results account for different values of $F_V$, $G_V$, and $F_A$: SD couplings from Eq.~\eqref{eq:par_SD} (upper half of each set of fits) and phenomenological couplings from Eq.~\eqref{eq:par_ph} (lower half of each set of fits). The uncertainties refer to the fit errors, prior to scale-factor inflation (where applicable).}
    \label{tab:fit_res}
\end{table}

Finally, for the computation of the uncertainty in the corrections to the HVP integral $a_\mu^\text{HVP, LO}[\pi\pi,\tau]$, the uncertainty due to the scheme dependence of $S_\text{EW}^{\pi\pi}$ as reported in Ref.~\cite{Aliberti:2025beg} is included. The same  uncertainty also affects the spectral function~\eqref{eq:Belle_spectrum}, but only via an overall rescaling. Accordingly, we first present fit results that focus on the uncertainty components listed above, while adding the SD error propagated from $S_\text{EW}^{\pi\pi}$ at the end.

\subsection{Fit results}

We perform fits for different degrees of the conformal polynomial ($N=3,4,5$, corresponding to $N-2$ free parameters) to the Belle data only, to Belle+ALEPH, and to all the data sets combined (Belle+ALEPH+CLEO+OPAL). The motivation to consider these combinations is as follows: first, the Belle data set provides the 
most precise spectral function, especially in the $\rho'$, $\rho''$ region, and we find that individual fits to the other data sets struggle to resolve the detailed resonance structure, which would require using a simplified fit function in these cases, and thus leading to results that are difficult to compare. For this reason, we always include the Belle data in each fit variant. Second, the only other data set for which a full documentation of the statistical and systematic uncertainties is available is ALEPH, motivating a combined fit with Belle. We still quote the result of the global fit to all data sets as our final result, but it is instructive to compare the outcome of the three fit variants as shown in    
Table~\ref{tab:fit_res}. 

In general, we obtain stable fits 
for $N=3,4$, with $p$-values ranging from a few percent if we consider only the Belle data set, over $3.0\times 10^{-3}$ for the fit to Belle+ALEPH, to $0.6\times 10^{-3}$ for the global fit. Accordingly, it is clear that some tensions are present in the data base, since the fit quality deteriorates considerably in the latter cases. To investigate the fit quality further, we also studied variants allowing for an additional parameter in the conformal polynomial, in which case the $p$-value increases to over $60\%$ for Belle only and to a few percent for the combined fit, suggesting that the fits with lower $N$ could be too constrained. However, we observe that the gain in the $\chi^2$ values comes at the expense of clear signs of overfitting, i.e., the $\rho''$ parameters need to be stabilized by imposing penalty functions, as otherwise mass and width parameter would run away to unphysical values. In addition, the asymptotic behavior of $\Im f_+(s)$ outside the region constrained by data deteriorates, indicating that tensions in the physical region are transferred to the high-energy tail. For that reason, we define our central values by $N=4$, while including the variation to $N=3,5$ in the uncertainty estimate. This procedure ensures that both the systematic variation among all fit variants and the error inflation of the statistical error are taken into account. The latter would be minimized or even absent if choosing the $N=5$ fits as central values, thus  hiding the tensions in the data base. 

\begin{figure}[t]
\centering
\includegraphics[width=0.8\linewidth]{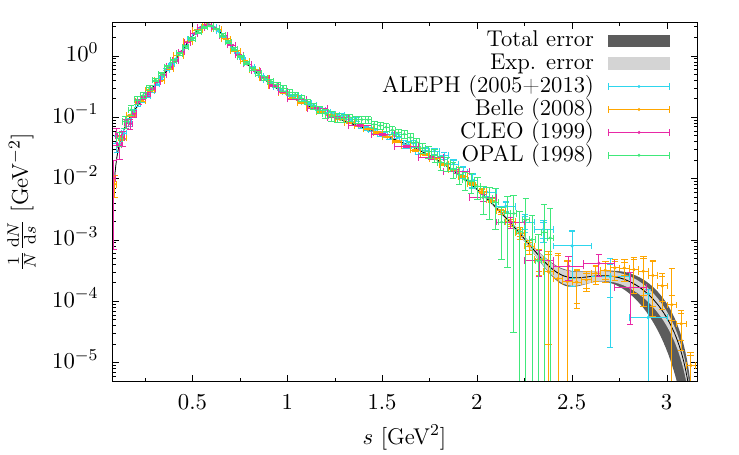}
\caption{Global fit to the $\tau\to\pi\pi\nu_\tau$ from Belle~\cite{Belle:2008xpe}, ALEPH~\cite{ALEPH:2005qgp,Davier:2013sfa}, CLEO~\cite{CLEO:1999dln}, and OPAL~\cite{OPAL:1998rrm}. Where available, inner error bars on the data points refer to statistical uncertainties and outer error bars to total ones. The ``Exp.\ error'' band on the curve corresponding to the global fit refers to the error coming from the uncertainty on the fit parameters inflated by a scale factor of $\sqrt{\chi^2/\text{dof}}$, see Table~\ref{tab:fit_res}, taking into account their correlations, as well as the propagated error of the ratio of branching fractions $\Br[\tau\to e\nu_\tau\bar{\nu}_e]/ \Br[\tau\to \pi\pi\nu_\tau]$ added in quadrature.}
    \label{fig:fullfit}
\end{figure}
\begin{figure}[t]
\centering
\begin{minipage}{0.49\linewidth}
    \centering
    \includegraphics[width=\linewidth]{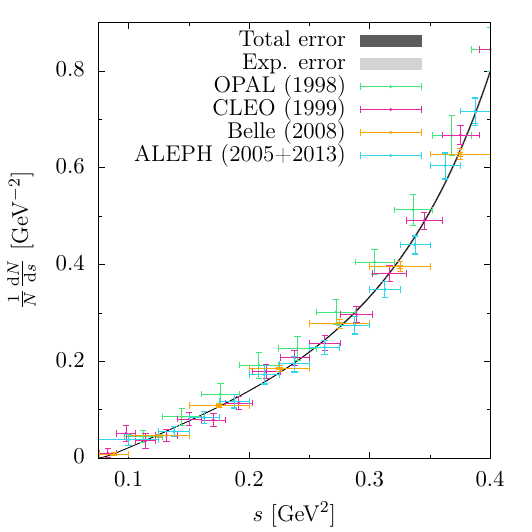}
\end{minipage}
\begin{minipage}{0.49\linewidth}
    \centering
    \includegraphics[width=\linewidth]{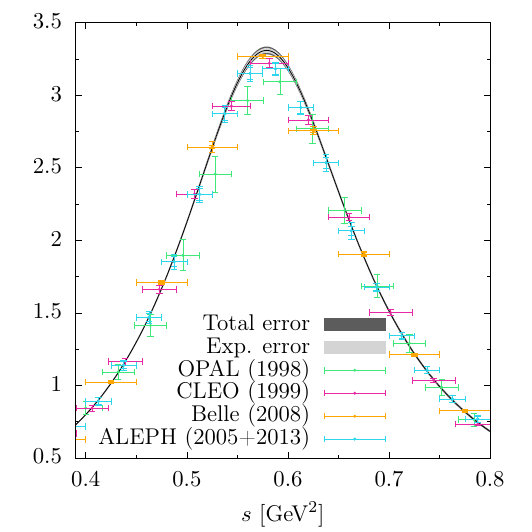}
\end{minipage}
\caption{Close-ups of the global fit to the spectrum in Fig.~\ref{fig:fullfit}. The low-energy region is shown in the left panel, while the $\rho(770)$ resonance region is shown in the right one.}
    \label{fig:fullfit_close}
\end{figure}

We also considered fit variants in which the explicit $\rho'$, $\rho''$ contributions are replaced by corresponding poles in the conformal variable~\cite{Kirk:2024oyl}. In general, the behavior is similar, as fits with smooth high-energy behavior tend to display relatively poor fit quality, which can be overcome by allowing for more fit parameters, but again at the expense of overfitting and even more severe instabilities in the $\rho'$, $\rho''$ parameters, presumably due to the fact that the sensitivity to the pole parameters is reduced compared to parameterizations better tailored for the real axis. While eventually the former would be preferred, we conclude that with the $\rho''$ so close to the border of the available phase space, fits to the $\tau$ spectral function based on the functional form~\eqref{eq:piFF} are better controlled. 

\begin{figure}[t]
\centering
\begin{minipage}{0.49\linewidth}
    \centering
    \includegraphics[width=\linewidth]{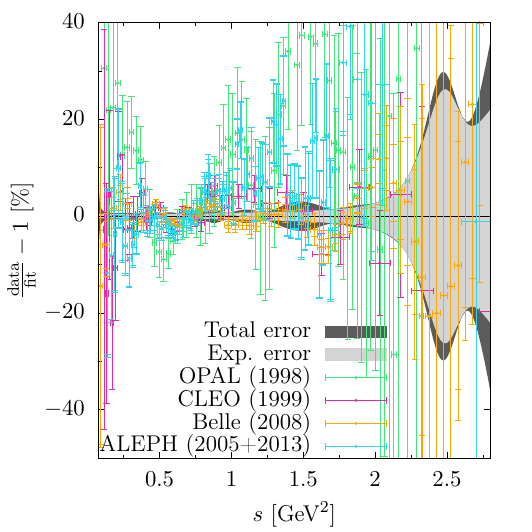}
\end{minipage}
\begin{minipage}{0.49\linewidth}
    \centering
    \includegraphics[width=\linewidth]{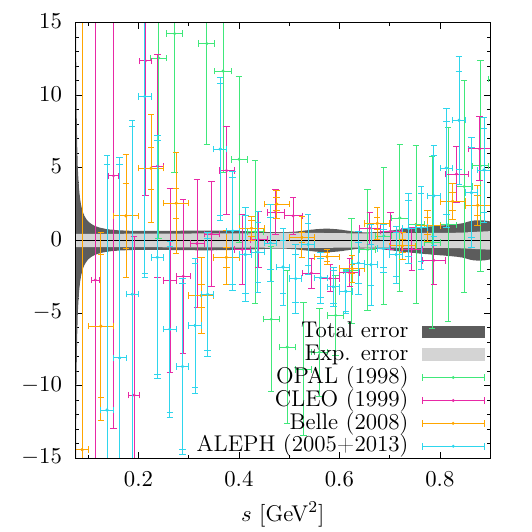}
\end{minipage}
\caption{Difference of the data relative to the global fit to the spectrum in Fig.~\ref{fig:fullfit}.}
    \label{fig:datafitdiff}
\end{figure}

\begin{figure}[t]
\centering
\includegraphics[width=0.8\linewidth]{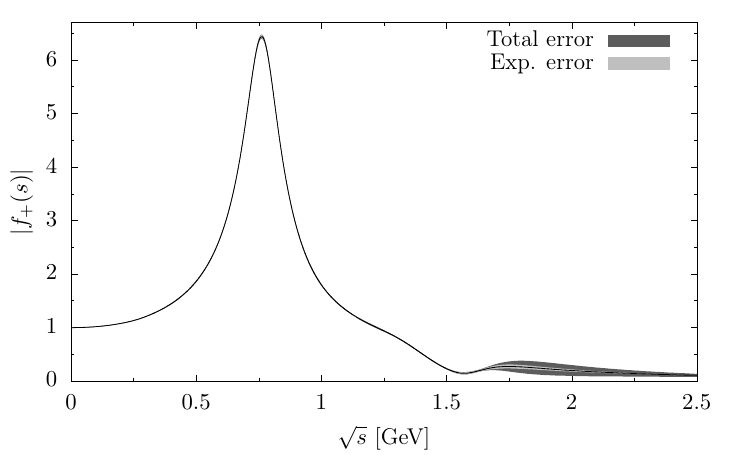}
\caption{Modulus of the pion VFF $|f_+(s)|$.}
    \label{fig:piVFF_ab}
\end{figure}

\begin{figure}[t]
\centering
\includegraphics[width=0.8\linewidth]{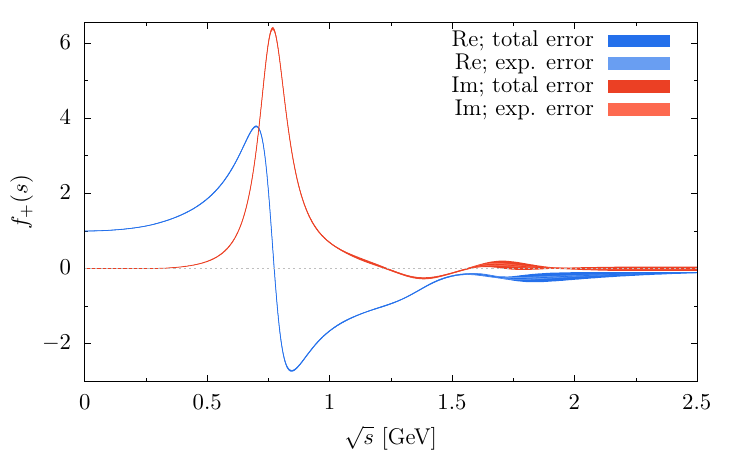}
\caption{Real and imaginary part of the pion VFF.}
    \label{fig:piVFF_reim}
\end{figure}

\begin{figure}[t]
\centering
\includegraphics[width=0.8\linewidth]{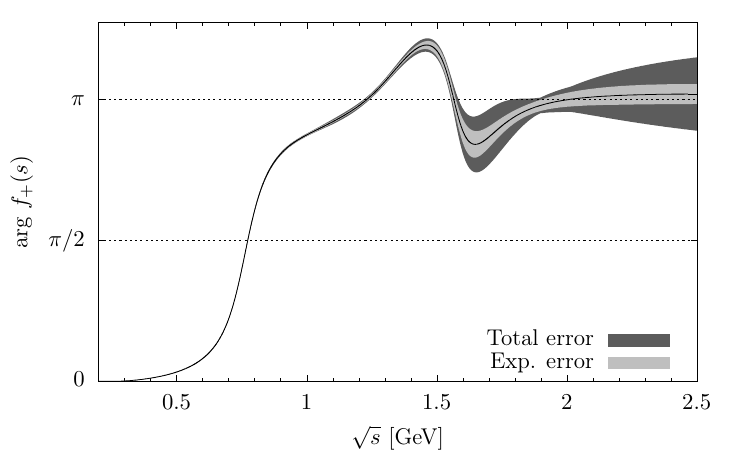}
\caption{Phase of the pion VFF.}
    \label{fig:piVFF_phase}
\end{figure}

The result of the global fit is shown in Fig.~\ref{fig:fullfit}, while in Fig.~\ref{fig:fullfit_close} we show close-ups of the global fit to the spectrum, for the low-energy region (left) and around the $\rho(770)$ resonance region (right). Additionally, in Fig.~\ref{fig:datafitdiff} we show the relative difference of the 
data to the global fit, for the full energy range (left) and a zoom around the region up to $0.9 \GeV^2$ (right). Due to the energy weighting in the calculation of $a_\mu^\text{HVP, LO}[\pi\pi,\tau]$, we find that the threshold region actually plays an integral role, especially, as there appears to be some (compensating) tension between the threshold and $\rho$-resonance region. That is, while in the central fits with $N=4$ the VFF in the $\rho$ peak tends to be overestimated, for $N=5$ the data around the $\rho(770)$ are better reproduced, yet the integrated $a_\mu$ value actually increases, due to an enhancement in the threshold region. Within uncertainties all fits are compatible, but the observation remains that the analyticity and unitarity constraints built into our dispersive representation of $f_+(s)$ suggest some tension in the data sets between threshold and resonance region.

The modulus of the pion VFF $f_+(s)$ arising from our global fit to the experimental data is shown in Fig.~\ref{fig:piVFF_ab}, while in Fig.~\ref{fig:piVFF_reim} we display the real and imaginary parts of $f_+(s)$ separately. 
It is also instructive to consider the phase of the pion VFF, 
see Fig.~\ref{fig:piVFF_phase}, especially in view of the preceding discussion about the asymptotic behavior. In all fit variants, by construction, the phase ultimately tends to $\pi$, ensuring the correct asymptotic behavior of the pion VFF $f_+(s)\simeq 1/s$~\cite{Chernyak:1977as,Farrar:1979aw,Efremov:1979qk,Lepage:1979zb,Lepage:1980fj}, but for $N=5$ one observes sizable oscillations before the phase returns to its asymptotic value.

\section{Consequences for \texorpdfstring{$\boldsymbol{a_\mu^\text{HVP, LO}[\pi\pi,\tau]}$}{}}
\label{sec:amu}

\begin{figure}[t]
\centering
\includegraphics[width=0.8\linewidth]{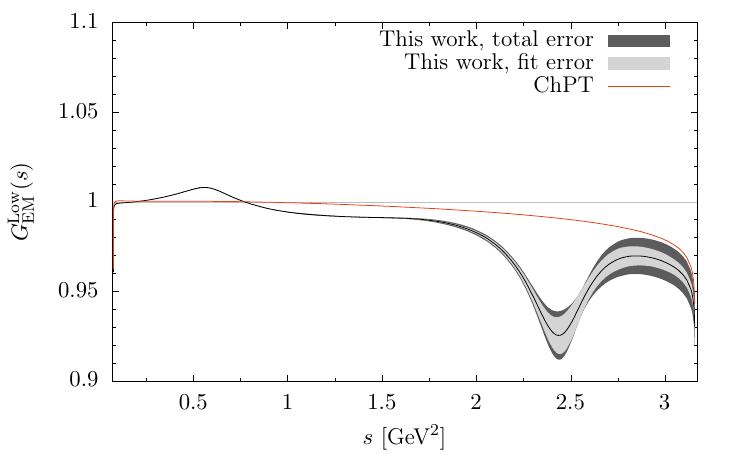}
\caption{Result for $G_\text{EM}(s)$ keeping only the leading Low contribution for real emission, $G_\text{EM}^\text{Low}(s)$. The ``fit error'' band refers to the error coming from the uncertainty propagated from the fit parameters of the form factor $f_+$ input inflated by a scale factor of $\sqrt{\chi^2/\text{dof}}$, see Table~\ref{tab:fit_res}, taking into account their correlations. In addition to our full dispersive evaluation, we also show the result using instead ChPT for the box diagram (based on Refs.~\cite{Cirigliano:2001er,Cirigliano:2002pv}).}
    \label{fig:gem_lead}
\end{figure}

\begin{figure}[t]
\centering
\includegraphics[width=0.8\linewidth]{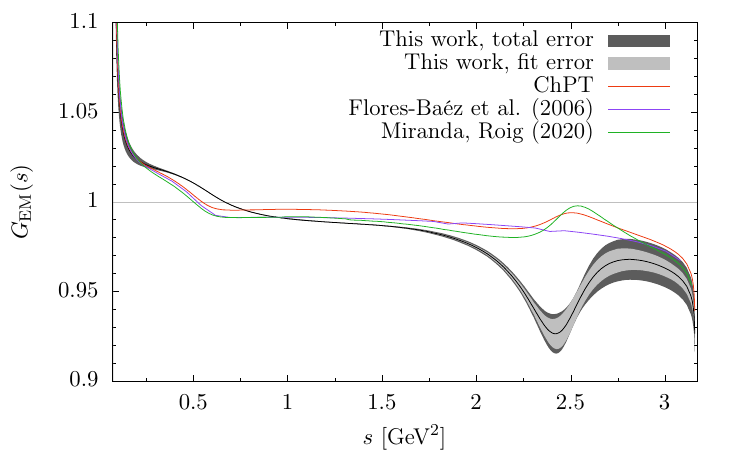}
\caption{Final result for $G_\text{EM}(s)$, compared to an implementation that uses ChPT for the evaluation of the box diagram (based on Refs.~\cite{Cirigliano:2001er,Cirigliano:2002pv}), as well as previous work by Flores-Ba\'ez et al.\ (2006)~\cite{Flores-Baez:2006yiq} and Miranda, Roig (2020)~\cite{Miranda:2020wdg}. Our numerical results are provided as supplementary material.}
    \label{fig:gem_full}
\end{figure}

Using the VFF fits presented in the previous section, we obtain the results for $G_\text{EM}(s)$ in Figs.~\ref{fig:gem_lead} and~\ref{fig:gem_full}. 
In particular, Fig.~\ref{fig:gem_lead} shows the result for the leading Low contribution to $G_\text{EM}(s)$, i.e., the one obtained by considering only the effect of $\tilde f_+^\text{match}(s,t)$ and $g_\text{Low}(s,t)$ in Eq.~\eqref{eq:delta_match}. Importantly, the curves labeled by ``ChPT'' in Figs.~\ref{fig:gem_lead} and~\ref{fig:gem_full} do not exactly correspond to the numerical results from Refs.~\cite{Cirigliano:2001er,Cirigliano:2002pv}, but are instead constructed replacing $\tilde f_+^\text{match}(s,t)$ in Eq.~\eqref{eq:delta_match} by $\tilde f_+^\text{ChPT}(s,t)$, which amounts to replacing the dispersive evaluation of the box diagram by its ChPT approximation. In this way, we obtain a more meaningful comparison of ChPT and dispersive results, updating other aspects of Refs.~\cite{Cirigliano:2001er,Cirigliano:2002pv} to the input used in this work, e.g., input for $f_+(s)$ and LECs.

The effect of the structure-dependent corrections in the $\rho$, $\rho'$, and $\rho''$ regions is clearly seen already in Fig.~\ref{fig:gem_lead}. While the ChPT curve remains almost constant with values around 1 for most of the energy range, our result becomes greater than $1$ around the $\rho$-resonance region, 
which affects the evaluation of the corresponding correction 
$\Delta a_\mu^\text{HVP, LO}[\pi\pi,\tau]$ due to $G_\text{EM}^\text{Low}(s)$ and leads to a substantial decrease by about $2.0\times 10^{-10}$. 
Figure~\ref{fig:gem_full} shows the full $G_\text{EM}(s)$, including also $g_\text{rest}(s,t)$ with all the resonance terms as detailed in Sec.~\ref{sec:ReEm}. While the qualitative behavior largely matches previous work, the energy dependence is altered substantially, most notably in the vicinity of the $\rho$ resonance. In addition, we observe a constant offset, which traces back to the local term in the ChPT contribution, especially the SD logarithm~\eqref{RG_LEFT}.\footnote{\label{footnote:local}Part of the local contribution to $f_+^\text{ChPT}(t)$, as given in Eq.~\eqref{CT+SE}, was absorbed into the definition of the pion VFF $f_+(s)$ in Refs.~\cite{Cirigliano:2001er,Cirigliano:2002pv}. We avoid this bookkeeping, as it would lead to a more complicated form of $f_+(s)$ in the fit to the $\tau$ spectral function, in particular, the normalization would differ from unity, by an amount ultimately controlled by the LEC $X_\ell$. Moreover, the analysis of the matching to SD contributions would become more complicated, given that this matching relation is most conveniently derived at the level of the chiral LEC~\cite{Cirigliano:2023fnz}.}

 \begin{table}[t]
    \centering
   \renewcommand{\arraystretch}{1.3}
	\begin{tabular}{lrrr}
	\toprule
	  & Belle & Belle+ALEPH & Belle+ALEPH+CLEO+OPAL\\\midrule
$G_\text{EM}^\text{Low}$ & $-2.292(15)(14)$ & $-2.279(13)(16)$ & $-2.267(13)(14)$\\
$G_\text{EM}^\text{rad}$ & $-5.21(3)(2)$ & $-5.20(3)(2)$ &$-5.19(3)(2)$ \\\midrule
$G_\text{EM}^\text{full}$ & $-5.44(3)(40)$ & $-5.43(3)(40)$ & $-5.41(3)(40)$\\
Phase space & $-7.74(4)(3)$ & $-7.73(4)(3)$ & $-7.74(4)(3)$\\
$S_\text{EW}^{\pi\pi}$ & $-12.180(57)(8)$ & $-12.177(57)(7)$ & $-12.166(56)(8)$\\
Sum & $-25.36(12)(44)$ & $-25.34(12)(44)$ & $-25.32(12)(44)$ \\\midrule
Full & $-24.84(12)(39)$ & $-24.82(12)(39)$ & $-24.80(12)(39)$\\\midrule 
$\tilde a_\mu$ & $510.1(2.4)(0.2)$ & $510.0(2.4)(0.2)$ & $509.5(2.4)(0.2)$\\
\bottomrule
	\renewcommand{\arraystretch}{1.0}
    \end{tabular}
    \caption{Result for the correction $\Delta a_\mu^\text{HVP, LO}[\pi\pi,\tau]$ (in units of $10^{-10}$) due to different $G_\text{EM}(s)$ contributions (leading Low, full radiation off $\tau$ and $\pi$, full real emission including resonances), phase-space factor, and $S_\text{EW}^{\pi\pi}$. The results labeled as ``Sum'' are obtained by summing the contributions from $G_\text{EM}^\text{full}(s)$, phase space, and $S_\text{EW}^{\pi\pi}$, while ``Full'' combines the same effects but without linearization. $\tilde a_\mu$ refers to the resulting two-pion contribution $a_\mu^\text{HVP, LO}[\pi\pi,\tau]$, but without consideration of IB in the matrix elements and before adding $e^+e^-$-specific corrections. For each entry, the errors refer to experimental and theory uncertainties, respectively, where the latter do not include yet the uncertainty due to the scheme ambiguity in $S_\text{EW}^{\pi\pi}$ nor an estimate of higher intermediate states in the virtual contribution. All results are provided separately for the three fit variants given in Table~\ref{tab:fit_res}.}
    \label{tab:IBres}
\end{table}

Given these results for $G_\text{EM}(s)$, we are now in the position to evaluate the shift of the HVP integral caused by IB corrections specific to the hadronic $\tau$ decay, $\tau^-\to\pi^-\pi^0\nu_\tau$. Setting $|F_\pi^V(s)/f_+(s)|=1$ in Eq.~\eqref{eq:RIB}, the full shift in $a_\mu^{\text{HVP, LO}}$ can be expressed as
\begin{equation}\label{eq:Deltaamu}
    \Delta a_\mu^{\text{HVP,LO}}[\pi\pi,\tau]=\bigg(\frac{\alpha m_\mu}{3\pi}\bigg)^2\int_{4M_\pi^2}^{m_\tau^2}\text{d}s\, \frac{\hat{K}(s)}{4s^2}\left[\frac{\big[\beta_{\pi\pi}(s)\big]^3}{\big[\beta_{\pi\pi^0}(s)\big]^3}\frac{1}{S_\text{EW}^{\pi\pi}G_\text{EM}[\tilde{s}(s)]}-1\right]v_\tau(s)\, ,
\end{equation}
with the $\tau$ spectral function given by
\begin{equation}
    v_\tau(s)=S_\text{EW}^{\pi\pi}\big[\beta_{\pi\pi^0}(s)\big]^3|f_+(s)|^2G_\text{EM}\big[\tilde{s}(s)\big]\, .
\end{equation}
Different contributions to $\Delta a_\mu^{\text{HVP,LO}}[\pi\pi,\tau]$ are usually singled out via linearization of the integrand
\begin{align}
    \Delta a_\mu^{\text{HVP,LO}}[\pi\pi,\tau]\bigg|_{r_\text{IB}}&=\bigg(\frac{\alpha m_\mu}{3\pi}\bigg)^2\int_{4M_\pi^2}^{m_\tau^2}\text{d}s\, \frac{\hat{K}(s)}{4s^2}\big[r_\text{IB}(s)-1\big]v_\tau(s)\, ,\notag\\
    r_\text{IB}(s)&=\Bigg\{\frac{\big[\beta_{\pi\pi}(s)\big]^3}{\big[\beta_{\pi\pi^0}(s)\big]^3}\,,
    \frac{1}{S_\text{EW}^{\pi\pi}}\,,
    \frac{1}{G_\text{EM}[\tilde{s}(s)]}\Bigg\}\,,
\label{eq:rIB}
\end{align}
to represent phase-space, SD, and radiative corrections, respectively, all of which encode different $\mathcal{O}(e^2)$ effects. Making this expansion explicit,  $M_{\pi^0}^2=M_\pi^2-\Delta_\pi$, $S_\text{EW}^{\pi\pi}=1+\Delta S_\text{EW}^{\pi\pi}$, and $G_{\text{EM}}(s)=1+\Delta G_\text{EM}(s)$, the difference of the linearized integrand and the full one, expanded in $e^2$, becomes
\begin{align}
    &\Bigg[\sum_{r_\text{IB}}\big[r_\text{IB}(s)-1\big]-\Bigg(\frac{\big[\beta_{\pi\pi}(s)\big]^3}{\big[\beta_{\pi\pi^0}(s)\big]^3}\frac{1}{S_\text{EW}^{\pi\pi}G_\text{EM}[\tilde{s}(s)]}-1\Bigg)\Bigg]v_\tau(s)\\
    &=-\frac{\sigma_\pi(s)}{s}\bigg[(s-4M_\pi^2)\Delta G_\text{EM}(s)\Delta S_\text{EW}^{\pi\pi}+3\Delta_\pi\big[\Delta G_\text{EM}(s)+\Delta S_{\text{EW}}^{\pi\pi}\big]\bigg]\big|f_+(s)\big|^2+\mathcal{O}\big(e^6\big)\, ,\notag
\end{align}
suggesting that since $G_\text{EM}(s)$ diverges at the $\pi\pi^0$ threshold, the term proportional to $\Delta_\pi\Delta G_\text{EM}(s)$ could become relevant due to the resulting threshold enhancement. 

The results for all these contributions are summarized in Table~\ref{tab:IBres}, further splitting the $G_\text{EM}(s)$ correction into the following contributions:
\begin{enumerate}
    \item $G_\text{EM}^\text{Low}(s)$ is obtained by setting to zero $g_\text{rest}(s,t)$ in Eq.~\eqref{eq:delta_match},
    \item $G_\text{EM}^\text{rad}(s)$ includes $g_\text{rest}(s,t)$, but with the resonance terms turned off, i.e., with $v_i$ and $a_i$ in Eq.~\eqref{eq:VmunuAmunu} and the $\omega$-interference contribution set to zero,
    \item $G_\text{EM}^\text{full}(s)$ includes the full resonance contribution as detailed in Sec.~\ref{sec:ReEm}.
\end{enumerate}
Finally, Table~\ref{tab:IBres} also includes a quantity defined as 
\beq
\tilde a_\mu\equiv \bigg(\frac{\alpha m_\mu}{3\pi}\bigg)^2\int_{4M_\pi^2}^{m_\tau^2}\text{d}s\, \frac{\hat{K}(s)}{4s^2}\frac{\big[\beta_{\pi\pi}(s)\big]^3}{\big[\beta_{\pi\pi^0}(s)\big]^3}\frac{v_\tau(s)}{S_\text{EW}^{\pi\pi}G_\text{EM}[\tilde{s}(s)]}\, ,
\eeq
which can be interpreted as the first step towards $a_\mu^{\text{HVP, LO}}[\pi\pi,\tau]$, prior to considering corrections in the matrix elements $|F_\pi^V(s)/f_+(s)|=1+\Order(e^2)$ and adding IB corrections specific for $e^+e^-$. 

\begin{table}[t]
    \centering
\renewcommand{\arraystretch}{1.3}
	\begin{tabular}{lrrrr}
	\toprule
& Ref.~\cite{Davier:2023fpl} & Ref.~\cite{Castro:2024prg} & Ref.~\cite{Aliberti:2025beg} & This work\\\midrule
Phase space & $-7.88$ & $-7.52$ & $-7.7(2)$ & $-7.74(5)$\\
$S_\text{EW}^{\pi\pi}$ & $-12.21(15)$ & $-12.16(15)$ & $-12.2(1.3)$ & $-12.2(1.3)$\\
$G_\text{EM}^\text{full}$ & $-1.92(90)$  & $(-1.67)^{+0.60}_{-1.39}$ & $-2.0(1.4)$ & $-5.4(5)$ \\
Sum & $-22.01(91)$ & $(-21.35)^{+0.62}_{-1.40}$ & $-21.9(1.9)$ & $-25.3(1.4)$ \\\midrule
Full & -- & -- & -- & $-24.8(1.4)$\\\midrule
$\tilde a_\mu$ & $510.3(3.0)$ & $510.9^{+2.9}_{-3.1}$ & $510.3(3.4)$ &$509.5(2.7)$\\
\bottomrule
	\renewcommand{\arraystretch}{1.0}
    \end{tabular}
    \caption{Comparison to previous work, following the notation of Table~\ref{tab:IBres}. Our results have been supplemented by the SD error from Ref.~\cite{Aliberti:2025beg}, which then dominates the final uncertainty, as well as an estimate of higher intermediate states in the virtual contribution.}
    \label{tab:IBcomp}
\end{table}

First, Table~\ref{tab:IBres} shows that the differences for the resulting IB corrections among the fits are small, which simply reflects the fact that the IB corrections are required with much less relative precision than the full integral. However, even at this level of precision we observe that a linearization of the IB corrections should be avoided, since the threshold-enhanced $\mathcal{O}(e^4)$ terms do become relevant. This observation also emphasizes the importance of a stable numerical implementation down to the two-pion threshold, to fully capture these corrections. 
Considering the changes among the different $G_\text{EM}(s)$ variants, the numerically largest contribution arises from the radiation off $\tau$ and $\pi$, around $-3.0\times 10^{-10}$, while the additional contribution due to resonance diagrams only induces an additional shift of $-0.2$ units. This shift, in fact, is less than half the size of the uncertainty propagated from the resonance couplings, most notably $F_A$, which dominates the overall uncertainty budget for $G_\text{EM}(s)$. In view of this substantial uncertainty already of the leading resonance contributions, which are motivated via ChPT resonance saturation, together with the overall small impact of resonance contributions on $\Delta a_\mu^\text{HVP, LO}[\pi\pi,\tau]$, we do not see a justification for including yet higher resonance multiplets. 

By comparing results for our full dispersive and the ChPT version of the box diagram, we find that structure-dependent virtual corrections amount to about $-2.0\times 10^{-10}$, yielding the second largest contribution after bremsstrahlung off $\tau$ and $\pi$.\footnote{The separation of real and virtual contributions is of course scale dependent, but the differences of dispersive and ChPT results for the box diagram, to quantify structure-dependent virtual contributions, and of $G_\text{EM}^\text{rad}$ and $G_\text{EM}^\text{Low}$, to quantify radiation off $\tau$ and $\pi$, are well defined.} Accordingly, one could worry about the possible impact of higher intermediate states in the hadronic matrix element, via resonance left-hand cuts or rescattering corrections. Given the experience from $\gamma^*\gamma^*\to\pi\pi$~\cite{Garcia-Martin:2010kyn,Hoferichter:2011wk,Moussallam:2013una,Danilkin:2018qfn,Hoferichter:2019nlq,Danilkin:2019opj}, one would expect such effects to be small in the low-energy region, with the first major resonance-enhanced structure related to the $f_2(1270)$ resonance. To account for virtual corrections beyond the pion pole, we assign an additional uncertainty of $0.3\times 10^{-10}$ to the $G_\text{EM}(s)$ contribution, motivated as the same relative uncertainty as resonance diagrams induce in the case of real emission.    

\begin{figure}[t]
\centering
\includegraphics[width=0.8\linewidth]{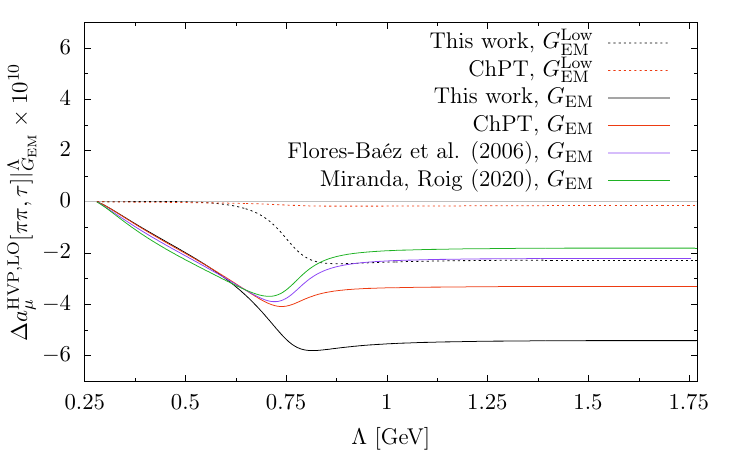}
\caption{$G_\text{EM}(s)$ contribution to $\Delta a_\mu^\text{HVP, LO}[\pi\pi,\tau]$ as a function of a cutoff $\Lambda$ in the HVP integral, for our dispersive implementation of the box diagram and its ChPT approximation, in both cases for the leading Low and full evaluation of real emission. Additionally, the results corresponding to using the full $G_\text{EM}(s)$ of previous works by Flores-Ba\'ez et al.\ (2006)~\cite{Flores-Baez:2006yiq} and Miranda, Roig (2020)~\cite{Miranda:2020wdg} are shown.}
    \label{fig:amu_gem}
\end{figure}

Our final results are summarized in Table~\ref{tab:IBcomp}, with central value defined by the $N=4$ VFF fits to all data sets and the mean of the strategies~\eqref{eq:par_SD} and~\eqref{eq:par_ph} for the resonance couplings
\beq
\label{Delta_amu_final}
\Delta a_\mu^\text{HVP, LO}[\pi\pi,\tau]\big|_\text{Full}=-24.8(0.1)_\text{exp}(0.5)_\text{th}(1.3)_\text{SD}\times 10^{-10}\,,
\eeq
where the experimental error combines the uncertainties derived from the covariance matrices of the data for the $\tau$ spectral function and the $\tau$ branching fractions, while the theory error accounts for all contributions listed in Sec.~\ref{sec:err}, i.e., variation of $N$, $s_c$, the cutoff of the dispersive integral, and the resonance couplings, with the latter the by far most important effect. Overall, the uncertainty is now dominated by the scheme dependence in $S_\text{EW}^{\pi\pi}$, that is, the matching between SD contributions contained in $S_\text{EW}^{\pi\pi}$ and radiative corrections described by $G_\text{EM}(s)$.  As for $G_\text{EM}(s)$ itself, however, the uncertainty has been reduced by almost a factor of three compared to the assignment in Ref.~\cite{Aliberti:2025beg}, which mainly reflects the fact that structure-dependent virtual corrections are now explicitly evaluated.    

It is also instructive to scrutinize the origin of the changes in central value compared to the previous work listed in Table~\ref{tab:IBcomp}, after all, our value for the $G_\text{EM}(s)$ contribution shifts by about $2.5\sigma$, part of which is then canceled upon adding the previously neglected $\Order(e^4)$ effects. To this end, we first evaluate the $G_\text{EM}(s)$ contribution as a function of the cutoff in the HVP integral, see Fig.~\ref{fig:amu_gem}, and compare the result to the ChPT evaluation of the box diagram, both for the leading Low and full calculation of the real-emission contributions. The figure shows that there are indeed significant differences in the energy dependence, as expected from Figs.~\ref{fig:gem_lead} and~\ref{fig:gem_full}, leading to the aforementioned decrease by $2.0\times 10^{-10}$ due to resonance enhancement of the $\rho(770)$.  Apart from this effect, further changes compared to Refs.~\cite{Aliberti:2025beg,Davier:2023fpl,Castro:2024prg} are surprising, as one would expect these evaluations to be closer to our ``ChPT'' result, but a large part of the difference traces back to the local contribution in Eq.~\eqref{CT+SE}, for which we use lattice-QCD-based input from Ref.~\cite{Ma:2021azh}.\footref{footnote:local}

\section{Summary and conclusions}
\label{sec:summary}

In this work we presented a comprehensive analysis of radiative corrections to $\tau\to\pi\pi\nu_\tau$ decays, including, for the first time, the effects of structure-dependent corrections in the loop integrals. To this end, we employed a dispersive representation of the pion VFF, capturing the dominant pion-pole contribution to the current--current matrix element that enters the unitarity diagram. We provided a detailed discussion how to implement the resulting correction numerically in a stable manner, how the dispersive representation reduces to its ChPT approximation in the appropriate limit, and how to match to ChPT by introducing a suitably chosen subtraction. We also revisited the real-emission diagrams, in particular, the numerical evaluation of the resulting correction close to the two-pion threshold, observing that the corresponding threshold enhancement requires the consideration of certain higher-order IB corrections. 

For the numerical evaluation of the IB corrections to the two-pion HVP integral derived from the $\tau\to\pi\pi\nu_\tau$ spectral function, self-consistent input for the pion VFF $f_+(s)$ is required. To this end, we performed fits to the available data sets, using a dispersive representation of $f_+(s)$ that implements constraints from analyticity, unitarity, and $\pi\pi$ Roy equations via the Omn\`es factor, supplemented by a conformal polynomial and explicit $\rho'$, $\rho''$ contributions to obtain a fit function that applies to the entire phase space. We observed that some tensions do exist in the data base, both among data sets and between the threshold and $\rho$-resonance region of the global fit, where the latter is a new feature that only becomes visible after imposing the general constraints from QCD. While the consistency of the data base is undoubtedly much better than for $e^+e^-\to\pi^+\pi^-$, these observations strongly motivate new high-statistics measurements of the $\tau\to\pi\pi\nu_\tau$ spectral function, as possible at  
Belle II~\cite{Belle-II:2018jsg}, which could also profit from the improved radiative-correction factor $G_\text{EM}(s)$ provided in this work. 

Finally, we evaluated the $\tau$-specific IB corrections to $a_\mu^\text{HVP, LO}[\pi\pi,\tau]$, derived from our results for $G_\text{EM}(s)$ and our VFF fits to the $\tau$ spectral function. For phase-space and short-distance corrections, we found good agreement with previous work, while for $G_\text{EM}(s)$ a larger negative correction was obtained, see Eq.~\eqref{Delta_amu_final} and Fig.~\ref{fig:gem_full} for our key results. We observed that changes due to structure-dependent contributions are indeed sizable in the vicinity of the $\rho(770)$, leading to a net correction of about $-2.0\times 10^{-10}$ in the HVP integral, while further changes to previous work trace back to the local ChPT contribution. 

A further main outcome of this work is the substantial reduction of the uncertainty in the $G_\text{EM}(s)$ contribution, leaving the matching between the short-distance factor $S_\text{EW}^{\pi\pi}$ and the radiative corrections described by $G_\text{EM}(s)$ as the dominant source of uncertainty in the $\tau$-specific IB corrections. This matching can be further improved using input from lattice QCD, and establishing the latter connection could also help address the remaining, most critical IB correction in the matrix elements. That is, our work allows for a reliable calculation of the long-range radiative corrections, but lattice-QCD techniques as well as complementary dispersive calculations of IB in the pion VFF, see Ref.~\cite{Colangelo:2025iuq} for a first step, are also needed to achieve a complete account of IB corrections to hadronic $\tau$ decays. Once all these ingredients become available, a robust evaluation of the two-pion HVP contribution to the anomalous magnetic moment of the muon will be possible.

\acknowledgments
We thank  Hisaki Hayashii, Bogdan Malaescu,  Lucas Mansur, Sven Menke, and Zhiqing Zhang for correspondence on various aspects of the $\tau\to\pi\pi\nu_\tau$ data sets. Further, we express our gratitude to Vincenzo Cirigliano, Christoph Greub, Pablo Roig, and Peter Stoffer for helpful discussions.
Financial support by the Swiss National Science Foundation (Project Nos.\ 200020\_200553 and TMCG-2\_213690) and the Albert Einstein Center for Fundamental Physics is gratefully acknowledged.


\appendix

\section{Phase-space integration for \texorpdfstring{$\boldsymbol{\tau^\pm\to\pi^\pm\pi^0\nu_\tau\gamma}$}{} }
\label{app:PhSpace}
In this appendix, we decompose the phase-space integration for the radiative $\tau$ decay into a convenient set of variables, and then discuss the physical regions to consider in order to compute the differential decay rate and spectral function. 

\subsection{Kinematics}

The kinematics of the process $\tau^-(l_1)\to\pi^-(q_1)\pi^0(q_2)\nu_\tau(l_2)\gamma(k)$ are described by the following invariants:
\begin{align}
    s&=(q_1+q_2)^2=(l_1-l_2-k)^2\,,\notag\\
    t&=(l_1-q_1)^2=(l_2+q_2+k)^2\,,\notag\\
    u&=(l_1-q_2)^2=(q_1+l_2+k)^2\,,\notag\\
    x&=(q_2+k)^2=(l_1-q_1-q_2)^2\,,\label{eq:kin2}
\end{align}
with $s+t+u+x=m_\tau^2+2M_\pi^2$, $l_1^2=m_\tau^2$, and $q_1^2=q_2^2=M_{\pi}^2$ (working at first order in IB, we again set $M_{\pi^0}=M_\pi$).
After the integration over neutrino and photon four-momenta the remaining integrals that need to be computed are $I_{mn}(s,t,x)$ given in Eq.~\eqref{eq:Imn}, which can be expressed as
\begin{equation}
    I_{mn}(s,t,x)=\frac{1}{2\pi}\int\frac{\text{d}^3l_2}{2l_2^0}\frac{\text{d}^3k}{2k^0}\delta\big(l_1^0-l_2^0-q_1^0-q_2^0-k^0\big)\frac{\delta^{(3)}\big(\lvec_1-\lvec_2-\qvec_1-\qvec_2-\kvec\big)}{(l_1\cdot k)^m(q_1\cdot k)^n}\,.
    \label{eq:Imn_ph}
\end{equation}
In the $\tau$ center-of-mass frame, the four-momenta read
\begin{align}
    l_1&=(m_\tau,\boldsymbol{0})\,,\qquad l_2=(|\lvec_2|,\lvec_2)\,,\qquad k=(|\kvec|,\kvec)\,,\notag\\
    q_1&=\left(\sqrt{M_\pi^2+|\qvec_1|^2},\qvec_1\right)\,,\qquad q_2=\left(\sqrt{M_\pi^2+|\qvec_2|^2},\qvec_2\right)\,.
\end{align}
By imposing three-momentum conservation in this reference frame, i.e., $\lvec_2=-\qvec_1-\qvec_2-\kvec$, and from the definition of the invariant $x$ in Eq.~\eqref{eq:kin2}, we get
\begin{equation}
    |\lvec_2|=\frac{1}{|\kvec|}\left(\frac{x}{2}-|\kvec|^2-|\kvec||\qvec_1+\qvec_2|\cos{\theta_k}\right)\,,
\end{equation}
with $\theta_k$ the angle between $\kvec$ and $\qvec_1+\qvec_2$. In these kinematics, energy conservation 
\begin{equation}
    f(|\kvec|)=m_\tau-|\lvec_2|-\sqrt{M_\pi^2+|\qvec_1|^2}-\sqrt{M_\pi^2+|\qvec_2|^2}-|\kvec|\equiv 0
\end{equation}
imposes 
\begin{equation}
    |\kvec|=|\kvec|_0=\frac{m_\tau x}{m_\tau^2-s+x+\cos\theta_k\sqrt{\lambda\left(s,x,m_\tau^2\right)}}\, ,
\end{equation}
and therefore
\begin{equation}
    I_{mn}=\frac{1}{2\pi}\int \text{d}\text{cos}\,\theta_k\;\text{d}\phi_k\; \text{d}|\kvec|\frac{|\kvec|^2}{4|\lvec_2||\kvec|}\frac{\delta\left(|\kvec|-|\kvec|_0\right)}{\left|f'\left(|\kvec|_0\right)\right|}\frac{1}{\left(l_1\cdot k\right)^m\left(q_1\cdot k\right)^n}\,,
\end{equation}
where
\begin{equation}
\label{Eq:meas}
    \frac{|\kvec|^2}{4|\lvec_2||\kvec|}\frac{1}{\left|f'\left(|\kvec|_0\right)\right|}=\frac{m_\tau^2 x}{2\left(m_\tau^2-s+x+\cos\theta_k\sqrt{\lambda\left(s,x,m_\tau^2\right)}\right)^2}\,.
\end{equation}

\begin{figure}[t]
\centering
\includegraphics[width=0.8\linewidth]{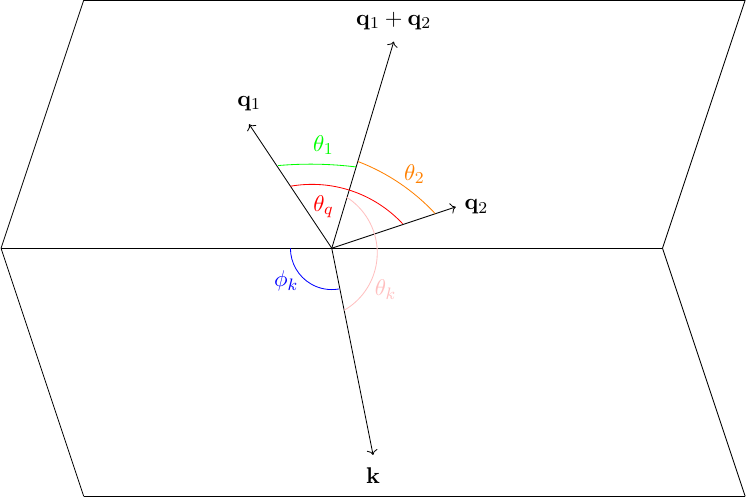}
\caption{The $\tau\to\pi\pi^0\nu_\tau\gamma$ decay in the $\tau$ rest frame.}
    \label{fig:angles}
\end{figure}

The remaining scalar products are decomposed as follows:
\begin{align}
 l_1\cdot q_1&=m_\tau\sqrt{M_\pi^2+|\qvec_1|^2}=\frac{1}{2}(m_\tau^2+M_\pi^2-t)\,,\quad |\qvec_1|=\frac{\sqrt{\lambda\left(t,m_\tau^2,M_\pi^2\right)}}{2m_\tau}\,,\notag\\
 l_1\cdot q_2&=m_\tau\sqrt{M_\pi^2+|\qvec_2|^2}=\frac{1}{2}(s+t-x-M_\pi^2)\,,\quad |\qvec_2|=\frac{\sqrt{\left(M_\pi^2-s-t+x\right)^2-4M_\pi^2m_\tau^2}}{2m_\tau}\,,\notag\\
 q_1\cdot q_2&=\frac{s-2M_\pi^2}{2}=\sqrt{M_\pi^2+|\qvec_1|^2}\sqrt{M_\pi^2+|\qvec_2|^2}-|\qvec_1||\qvec_2|\cos\theta_q\,,\notag\\
 &\qquad |\qvec_1+\qvec_2|=\sqrt{|\qvec_1|^2+|\qvec_2|^2+2|\qvec_1||\qvec_2|\cos\theta_q}\,,\notag\\
 l_1\cdot k &=m_\tau |\kvec|_0\, ,\notag \\
 q_1\cdot k&=|\kvec|_0\left(\sqrt{M_\pi^2+|\qvec_1|^2}-|\qvec_1|\cos\theta_{kq_1}\right)\, ,
 \label{scalar_products}
\end{align}
with $\theta_{kq_1}$ the angle between $\kvec$ and $\qvec_1$. By choosing the axes such that $\qvec_1$, $\qvec_2$, and accordingly $\qvec_1+\qvec_2$ lie in one plane, as shown in Fig.~\ref{fig:angles}, the following explicit choices can be made for the three-momenta,
\begin{equation}
    \qvec_1+\qvec_2=\begin{pmatrix}
         0 \\
         0 \\
         |\qvec_1+\qvec_2|
    \end{pmatrix}\,,\qquad 
\kvec=|\kvec|\begin{pmatrix}
         \sin\theta_k\cos\phi_k\\
         \sin\theta_k\sin\phi_k\\
         \cos\theta_k
    \end{pmatrix}\,,
\end{equation}
and
\begin{equation}
    \qvec_1=|\qvec_1|\begin{pmatrix}
         \sin\theta_1\\
         0\\
         \cos\theta_1
    \end{pmatrix}\,,\qquad 
    \qvec_2=|\qvec_2|\begin{pmatrix}
         -\sin\theta_2\\
         0\\
         \cos\theta_2
    \end{pmatrix}\,.
\end{equation}
Then, the angle $\theta_{kq_1}$ can be determined by
\begin{align}
\cos\theta_{kq_1}&=\sqrt{1-\cos^2\theta_k}\sqrt{1-\cos^2\theta_1}\cos\phi_k+\cos\theta_k\cos\theta_1\,,\notag\\
\cos\theta_1&=\frac{|\qvec_1|+|\qvec_2|\cos\theta_q}{\sqrt{|\qvec_1|^2+|\qvec_2|^2+2|\qvec_1||\qvec_2|\cos\theta_q}}\,,
\end{align}
where $\theta_q$ is given in Eq.~\eqref{scalar_products} as the angle between $\qvec_1$ and $\qvec_2$.
In this way, the integrand in $I_{mn}$ is expressed via masses, Mandelstam variables, and angles $\theta_k$ and $\phi_k$. The same kinematic reference frame can be used in the numerical phase-space integration of the full radiative amplitude given in Eq.~\eqref{eq:amp_ReEm}.

\subsection{Integration region and boundaries}

In order to calculate differential rates and spectra, we need the physical region $\mathcal{D}$ in the form of a normal domain:
\begin{equation}
    \mathcal{D}=\left\{s_\text{min}\leq s\leq s_\text{max},\:t_\text{min}(s)\leq t\leq t_\text{max}(s),\:x_\text{min}(s,t)\leq x\leq x_\text{max}(s,t)\right\}\,,
\end{equation}
where $s_\text{min}=4M_\pi^2$ and $s_\text{max}=m_\tau^2$. Following closely the notation of Ref.~\cite{Cirigliano:2002pv},  
the explicit form of $t_\text{min/max}(s)$ and $x_\text{min/max}(s,t)$ is conveniently expressed in terms of the functions
\begin{align}
    t_\pm(s,x)&=\frac{1}{2}\left[m_\tau^2+2\mpi^2-s+x\pm\sigma_\pi(s)\lambda^{1/2}\left(m_\tau^2,s,x\right)\right]\,,\notag\\
    x_\pm(s,t)&=\frac{1}{2M_\pi^2}\left[2M_\pi^2\left(m_\tau^2+s\right)-s\left(m_\tau^2+M_\pi^2-t\right)\pm s\sigma_\pi(s)\lambda^{1/2}_{\pi\tau}(t)\right]\,,
\end{align}
with $\lambda_{\pi\tau}(t)$ defined in Eq.~\eqref{Eq:B0_C0_lambda}.
\begin{figure}
    \centering
    \includegraphics[width=0.65\linewidth]{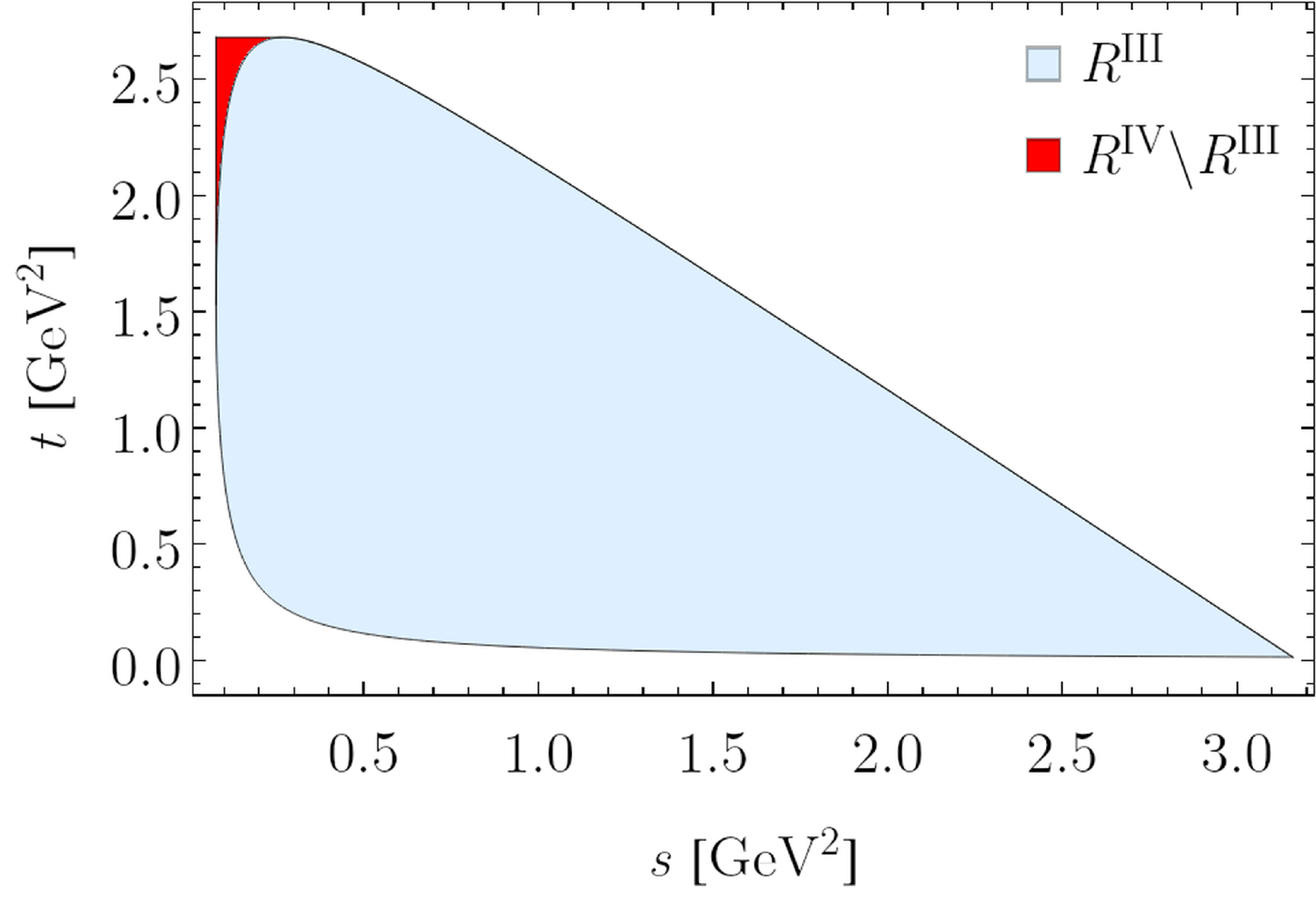}
    \caption{$(s,t)$ Dalitz plot, where $R^\text{IV}$ represents the full radiative phase space, while $R^\text{III}$ is accessible to the nonradiative decay. The contribution in  $R^\text{IV}\backslash R^\text{III}$ is important for the threshold behavior of $G_\text{EM}(s)$.}
    \label{fig:dalitz}
\end{figure}

Let us call $R^\text{III}$ the region in the $s$--$t$ plane accessible in the nonradiative three-body decay, and $R^\text{IV}$ the region accessible in the radiative decay, see Fig.~\ref{fig:dalitz}.
The boundary of
$R^\text{III}$ is given by
\beq
    \bar{t}_\text{min}(s)\equiv t_-(s,0)\,,\qquad 
    \bar{t}_\text{max}(s)\equiv t_+(s,0)\,,\qquad 
    s_\text{min}\leq s\leq s_\text{max}\,,
\eeq 
while $R^\text{IV}$ corresponds to
\begin{align}
    t_\text{min}(s)&\equiv t_-(s,0)\qquad \text{for all}\quad s_\text{min}\leq s\leq s_\text{max}\,,\notag\\
    t_\text{max}(s)&\equiv \begin{cases}
      t_+(s,0)\qquad  &\text{for}\quad s_*\leq s\leq s_\text{max}\, ,  \\
      \left(m_\tau-M_\pi\right)^2\qquad&\text{for}\quad s_\text{min}\leq s\leq s_*\, ,
    \end{cases} \qquad 
 s_*\equiv \frac{m_\tau^2 M_\pi}{m_\tau-M_\pi}\,.   
\end{align}
In each case, for given $(s,t)$, the limits for $x$ are
\begin{align}
    x_\text{max}(s,t)&\equiv x_+(s,t)\,,\notag\\
    x_\text{min}(s,t)&\equiv \begin{cases}
         0 & \text{for}\quad (s,t)\in R^\text{III}\, ,\\
         x_-(s,t)&\text{for}\quad(s,t)\in\ R^\text{IV}\backslash R^\text{III}\, .
    \end{cases}
\end{align}
Finally, in Eq.~\eqref{eq:Js} we defined the quantities
\begin{equation}
    Y_{1,2}=\frac{1-2\bar{\alpha}\pm\sqrt{\left(1-2\bar{\alpha}\right)^2-\left(1-\bar{\beta}^2\right)}}{1+\bar{\beta}}\,,
\end{equation}
with
\begin{align}
    \bar{\alpha}&=\frac{\left(m_\tau^2-s\right)\left(m_\tau^2+M_{\pi}^2-s-t\right)}{M_\pi^2+m_\tau^2-t}\times\frac{\lambda_{\pi\tau}(t)}{2\bar{\delta}}\,,\qquad 
    \bar{\beta}=-\frac{\lambda_{\pi\tau}^{1/2}(t)}{M_\pi^2+m_\tau^2-t}\,,\qquad 
    \bar{\gamma}=\frac{\lambda_{\pi\tau}^{1/2}(t)}{2\sqrt{\bar{\delta}}}\,,\notag\\
    \bar\delta&=st(m_\tau^2+2\mpi^2-s-t)-\mpi^2 m_\tau^2(m_\tau^2-s)-\mpi^4 s\,.
\end{align}

\bibliographystyle{apsrev4-1_mod_2}
\bibliography{amu}

\end{document}